\documentclass[aps,prb,twocolumn,amsmath,amssymb]{revtex4}
\usepackage{graphicx}
\usepackage{dcolumn}
\usepackage{bm}
\usepackage{color}
\definecolor{DarkBlue}{rgb}{0.1,0.1,0.5}
\definecolor{Red}{rgb}{0.9,0.0,0.1}
\definecolor{Green}{rgb}{0.0,0.99,0.0}

\begin{document}

\title{Orbital degeneracy as a source of frustration in LiNiO$_2$}

\author{F. Vernay$^1$, K. Penc$^2$, P. Fazekas$^2$, F. Mila$^1$}

\affiliation{
$^1$Institute of Theoretical Physics, Ecole Polytechnique 
F\'ed\'erale de Lausanne, CH-1015 Lausanne, Switzerland\\
$^2$Research Institute for Solid State 
Physics and Optics, H-1525 Budapest, P.O.B. 49, Hungary. }

\date{\today}

\begin{abstract}
Motivated by the absence of cooperative Jahn-Teller effect 
and of magnetic ordering in
LiNiO$_2$, a layered oxide with triangular 
planes, we study a general spin-orbital model on the 
triangular lattice. A mean-field approach reveals the presence
of several singlet phases between the SU(4) symmetric point
and a ferromagnetic phase, a conclusion supported by exact 
diagonalizations of finite clusters. We argue that one of the 
phases, characterized by a large number of low-lying singlets 
associated to dimer coverings of the triangular lattice,
could explain the properties of LiNiO$_2$, while a ferro-orbital
phase that lies nearby in parameter space leads to a new
prediction for the magnetic properties of NaNiO$_2$.
\end{abstract}

\maketitle

\section{Introduction}
The Mott insulators LiNiO$_2$ and NaNiO$_2$ are isostructural and 
isoelectronic, but they have completely different phase diagrams. The 
complicated nature of these systems arises from an interplay of the 
dynamical frustration of spin--orbital models with the geometrical frustration 
of the triangular lattice which is the essential structural unit. We will show 
that by a modest change of parameters, a great variety of phases can be 
derived.

\begin{figure}[ht]
\begin{center}
\includegraphics*[width=7truecm,angle=0]{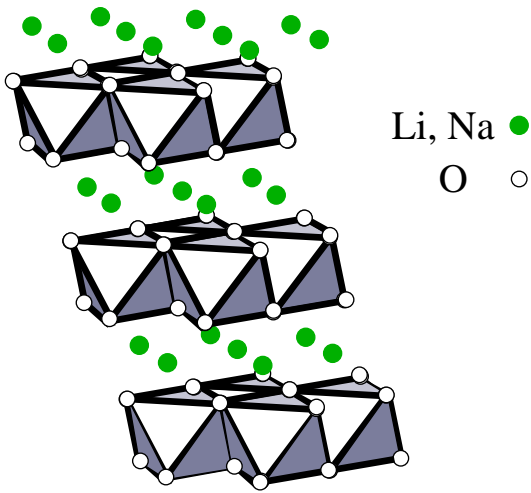}
\caption[ANiO2]{\label{ANiO2} 
ANiO$_2$ structure. Ni ions are located in the middle of the O octahedra.}
\end{center}
\end{figure}

The crystal structure can be envisaged as a sequence of slabs of edge sharing 
octahedra of oxygen O$^{2-}$ ions. Metal ions sit at the centers of
octahedra. There are two kinds of slabs: in A slabs, at every center of 
octahedra there is a Ni$^{3+}$, whereas in the B slabs, one finds either 
Li$^+$ or Na$^+$ ions. A and B slabs alternate (see Fig.~\ref{ANiO2}). 
The Ni ions form well-separated triangular planes.  

It is useful to start with the idealized geometry of a cubic
system. Neglecting the inequivalence of Ni and Li sites, and assuming 
perfect oxygen octahedra, the octahedral centres would form a simple cubic
lattice. The slabs of the original structure would be perpendicular to the 111
direction. Within a slab the Ni--O--Ni bond angles would be 90$\,^{\circ}$,  
resulting in important consequences for the effective exchange \cite{most}.  

There are two sources of deviation from cubic symmetry: a) Ni and Li/Na sites
are inequivalent, which leaves us with one (instead of four) ${\cal C}_3$
axis. Even if the octahedra were undistorted, Ni ions would see a wider 
environment with trigonal symmetry only. b) actually, oxygen octahedra 
are distorted\cite{chap-JT,yamaura}, and the  Ni--O--Ni bond angle is 
$\approx 96.4\,^{\circ}$ in the case of Na, and $\approx 94\,^{\circ}$
 in case of the Li compound.

If there is a Jahn--Teller phase transition (as in NaNiO$_2$), it lowers the 
crystal symmetry further, and makes the orbital ground state unique. An 
alternative would be to ascribe orbital polarization to an electronic 
phase transition due to orbital exchange, and to regard the lattice 
distortion as an induced secondary effect. In what follows, we assume 
trigonal point group symmetry which is valid for NaNiO$_2$ at high 
temperatures, and for LiNiO$_2$ at all temperatures. Breaking the local 
trigonal symmetry, whenever it happens, is ascribed to orbital ordering. We 
consider electronic degrees of freedom only, but we assume that the 
lattice would follow the changing orbital occupation.   
 
The Ni$^{3+}$ ions are in the $S=1/2$ low-spin state. In terms of the 
dominant cubic component of the crystal field $3d^7=t_{2g}^6e_g^1$. Since the 
actual point group symmetry is trigonal, $t_{2g}$ gets split into two levels 
($t_{2g} \rightarrow A_2 + E$, where standard notations for the irreps of 
the point group ${\cal D}_{3d}$ were introduced)
but this does not affect the fact that 6 electrons are taken up by closed 
subshells, and only the seventh electron is in an open subshell. The trigonal 
crystal field component 
changes the detailed nature of the $d$-states, but still allows for 
twofold orbital degeneracy: $e_g^{\rm cubic}\rightarrow  E$. 
In what follows, $E$ is understood to denote the two-dimensional irrep of 
the trigonal point group\cite{otherE}.

The ground state of an isolated Ni$^{3+}$ ion is fourfold degenerate: it has 
twofold orbital, and twofold spin degeneracy. A standard scenario would be 
that the non-Kramers degeneracy is resolved by a (cooperative) Jahn--Teller 
effect, while the Kramers degeneracy is lifted by magnetic ordering. Let us 
note that, as far as the $E$-electrons are concerned,  
the cooperative Jahn--Teller effect is synonymous with orbital ordering, 
thus it can be explained with a purely electronic model, without the 
consideration of electron--lattice coupling.

The standard scenario seems to be (at least nearly) realized for NaNiO$_2$, 
which has a first order cooperative Jahn-Teller transition lowering the local 
symmetry from trigonal to monoclinic at $T_{\rm JT}^{\rm Na}\sim 480{\rm K}$ 
\cite{chap-JT}. The 
remaining Kramers degeneracy is lifted by a magnetic transition at 
$T_{\rm N}^{\rm Na}\approx 20{\rm K}$, which was characterized as the 
antiferromagnetic ordering of ferromagnetic planes \cite{chap-mag}.   
In contrast, LiNiO$_2$ does not undergo a Jahn--Teller 
distortion\cite{rougier}, and 
though the measured susceptibility shows a number of anomalies, it does not 
seem to develop magnetic long range order \cite{hirota,yamaura,kitaoka}. 
It is puzzling that the two isostructural and isoelectronic compounds 
show so different behavior. Naturally, it is always possible that some 
of the observed behavior is not intrinsic. Impurities and structural 
defects are likely to prevent orbital ordering. Indeed, it was suggested that 
only NaNiO$_2$ allows the growth of sufficiently good-quality samples, 
and the observation of ordering transitions, while the overall behavior of 
LiNiO$_2$ is like that of the high-temperature phase of NaNiO$_2$ 
\cite{chap-JT}. 

\section{Spin--orbital model based on the trigonal doublet}

The aim of the present paper is to show that the contrasting features 
of NaNiO$_2$ and LiNiO$_2$ appear naturally as nearly equivalent  
possibilities for the intrinsic behavior of spin-orbital models of the 
trigonal $E$ doublet.   

A similar four-state model, namely the $S=1/2$ cubic $e_g$ doublet on the 
cubic lattice, has been studied in great detail in the context of manganite 
physics\cite{olesrev}. The magnetic behaviour is complicated because 
orbital and spin--orbital interactions tend to frustrate the usual spin--spin
interactions. Though, in contrast to spin-only models, spin--orbital models 
do not need the fine-tuning of the lattice structure to get frustration 
effects, we find that the geometrical frustration of the triangular planes 
of the LiNiO$_2$ structure brings essential new features. For this reason, 
we consider only an isolated triangular plane, and discuss $T=0$ behavior 
only. We assume that our essential conclusions would carry over to the $T>0$ 
behavior of coupled planes. 

The idea that the geometrical frustration of the triangular lattice tends to 
oppose ordering, has been discussed for spin\cite{rvb,kleine} and 
orbital\cite{rey} degrees of freedom separately. In a pioneering work, 
Hirakawa et al.\cite{hira} started a systematic investigation of triangular 
lattice antiferromagnets with the explicit aim of finding non-N\'eel-type 
behavior. This work initiated the intensive re-investigation of LiNiO$_2$. 
On the theoretical side, Arimori and Miyashita\cite{arim} studied a 
classical model and found that novel order parameters combining 
spin and orbital character are important. In a quantum-mechanical 
calculation, chosing a special set of parameters to make the four-state 
spin--orbital model SU(4) symmetrical, it was found that the ground 
state of the 
nearest-neighbour model on the triangular lattice is an SU(4)-resonating 
quantum liquid\cite{su4penc}.  

Here we consider the full range of $E$ models, restricting the parameters
only by the requirements dictated by symmetry. The pair interaction is 
generically of SU(2)${\otimes}{\cal C}_{2h}$ symmetry; higher symmetries 
(SU(2)${\otimes}$SU(2) or SU(4)) follow from specific choices of the 
parameters. We explore many lower-symmetry states in addition 
to the fully symmetrical SU(4) phase.  
Accepting that the observed behavior of both LiNiO$_2$ and NaNiO$_2$ is 
intrinsic, any theory for why LiNiO$_2$ does not order should also allow for 
the alternative scenario of orbital and spin ordering, as observed in 
NaNiO$_2$. In terms of our trigonal $E$ model, 
we show that the combination of geometrical frustration with
the dynamical frustration inherent in spin-orbital models gives rise to a rich
variety of competing states stretching from the SU(4) resonating singlet 
state to spin-ferromagnetic phases with various orbital order. We will find it
natural that contrasting behavior resembling that of either LiNiO$_2$ or 
NaNiO$_2$ can arise in nearby regions of parameter space.

\subsection{Basis functions}

Our model is meant to describe the Mott-localized $E$ electrons of Ni ions. 
The local degrees of freedom are those of an $E^1$ shell. Intersite 
interactions arise from the virtual charge fluctuations 
$E^1E^1\rightarrow E^2E^0$. The study of such spin--orbital exchange 
models was initiated by Kugel and Khomskii\cite{KK}, and by Castellani 
et al.\cite{cast}.   

The point group of a Ni site is ${\cal D}_{3d}={\cal D}_3{\otimes}\{{\cal E},
{\cal I}\}$, where ${\cal E}$ is the identity element, and ${\cal I}$ is the 
inversion. The subgroup of proper rotations ${\cal D}_3$ contains the trigonal 
axis ${\cal C}_3$ 
and three orthogonal ${\cal C}_2$ axes. 
It is convenient to denote axes in terms of the original octahedral 
system $\{ X,Y,Z\}$, so the ${\cal C}_3$ axis is (111). For later reference, 
we recall that ${\cal D}_{3d}$ has three irreps: the identity rep $A_1$, the 
one-dim irrep $A_2$, and the two-dim irrep $E$.

\begin{figure}[ht]
\begin{center}
\includegraphics*[width=2.5truecm,angle=0]{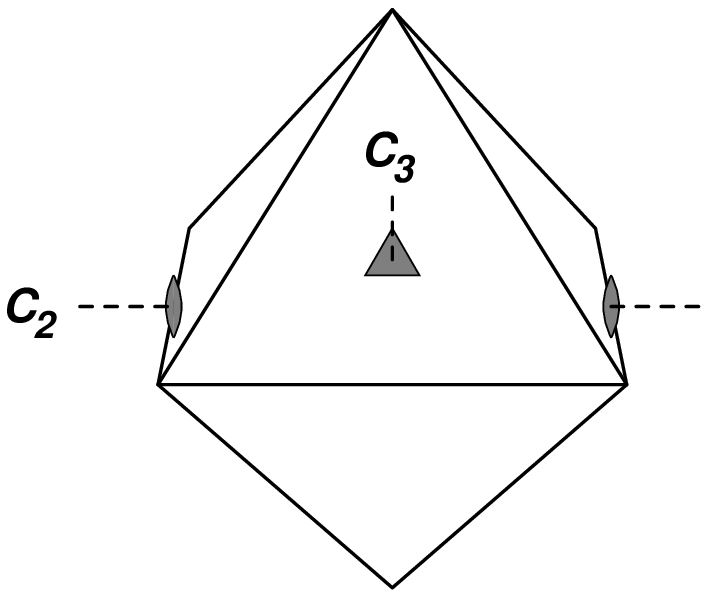}
\hspace{0.5cm}
\includegraphics*[width=4truecm,angle=0]{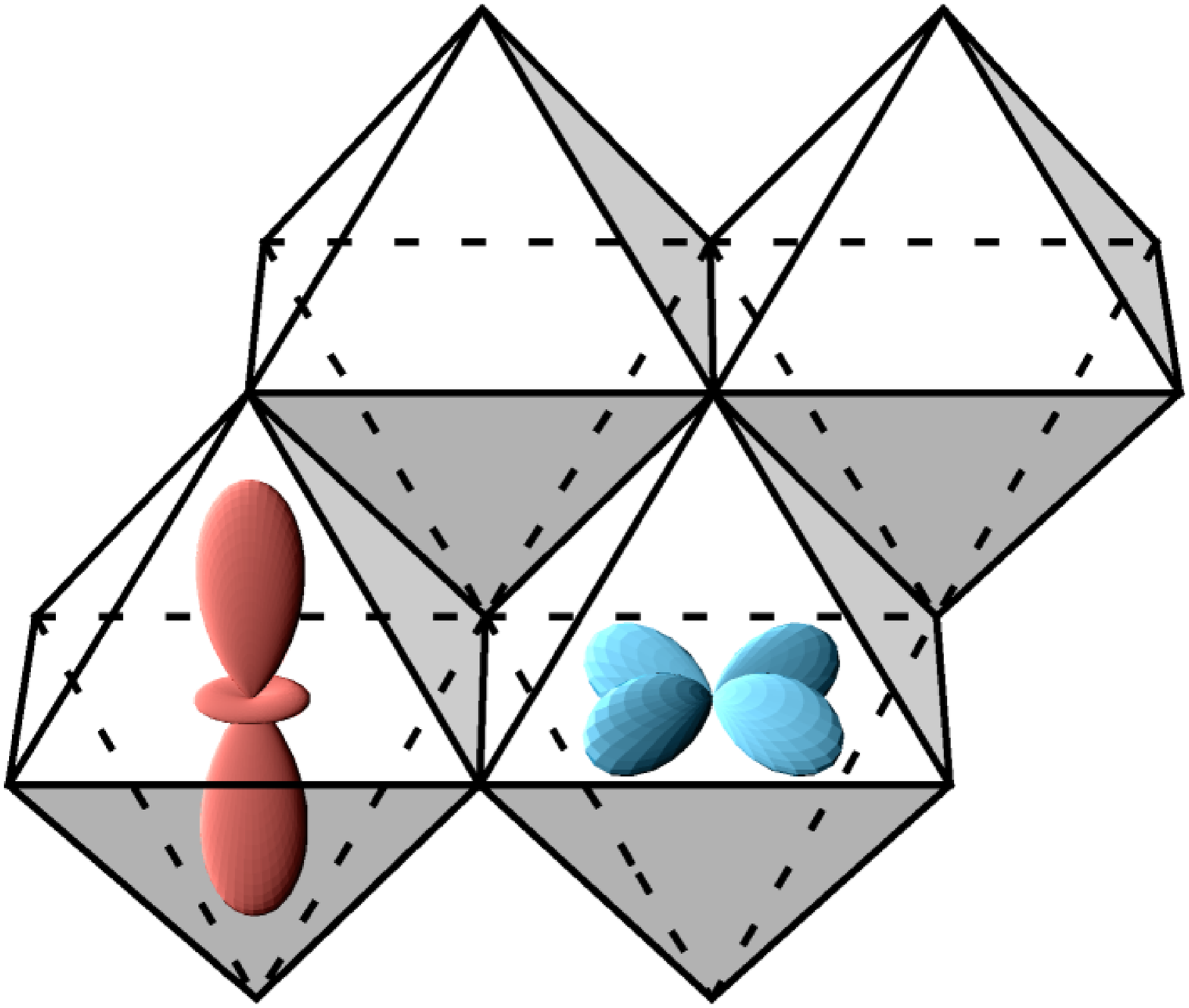}
\caption[niv]{a. The ${\cal C}_3$ axis and one of the ${\cal C}_2$ axes of the 
point group of the Ni site in the ANiO$_2$ structure (the other two 
${\cal C}_2$ axes are obtained by applying ${\cal C}_3$). b. Orbital states 
of the seventh $d$-electron of Ni$^{3+}$ on the backgound of the network of 
oxygen octahedra. 
\label{orbs}}
\end{center}
\end{figure}

First, we represent ${\cal D}_3$ on the basis of the $E$ subspace spanned by 
$c^{\dagger}_a|0\rangle=|a\rangle\propto (3Z^2-R^2)$, 
and $=c^{\dagger}_b|0\rangle=|b\rangle\propto (X^2-Y^2)$ (as yet, we omit 
the spin index). Alternatively, we may represent on the 
two-dimensional operator 
subspaces $\{c^{\dagger}_a, c^{\dagger}_b\}$ (or $\{c_a, c_b\}$). 
The effect of a $\pi$-rotation about the $1{\overline 1}0$ axis (one of the 
${\cal C}_2$ axes) is
\begin{eqnarray}
c^{\dagger}_{a'} = c^{\dagger}_{a'}
{\mbox{\ \ \ \ \ \ \ \ \ \ }} c^{\dagger}_{b'} = -c^{\dagger}_{b'}
\, .
\label{eq:trans1}
\end{eqnarray}
Skipping the effect of the other two ${\cal C}_2$ rotations, we 
show how a $2\pi/3$ rotation about the trigonal axis is represented in the 
 $E$ subspace 
\begin{eqnarray}
  c^{\dagger}_{a'}&=& -\frac{1}{2} c^{\dagger}_{a} +
  \frac{\sqrt{3}}{2}   c^{\dagger}_{b} \nonumber\\
  c^{\dagger}_{b'}&=& -\frac{\sqrt{3}}{2} 
c^{\dagger}_{a} -\frac{1}{2} c^{\dagger}_{b}\, .
\label{eq:trans2}
\end{eqnarray}

\subsection{The microscopic model}

On-site $d$-electron orbital states are classified according to the point 
group ${\cal D}_{3d}$ \cite{double}, while two-site states according to 
the smaller point group ${\cal C}_{2h}$ of a pair.   
The nearest neighbours of a Ni site are at the centers of octahedra which 
share an edge with the first site. The ${\cal C}_2$ axis perpendicular to 
this edge is a symmetry element of the pair; so is the mirror plane 
$\sigma_h$ perpendicular to the ${\cal C}_2$ axis in question\cite{inver}.  
${\cal C}_2$ and $\sigma_h$ are the generators of the 4-element symmetry 
group ${\cal C}_{2h}$ of the pair\cite{lattice}.

The standard components of the electronic hamiltonian are those of a 
two-band extended Hubbard model: intersite hopping ${\cal H}_{\rm hop}$, and 
on-site Coulomb matrix elements ${\cal H}_{\rm Coul}$.

First, we discuss ${\cal H}_{\rm hop}$. The local $E$ basis can always be 
chosen so that under the ${\cal C}_2$ rotation of the pair, one of the basis 
states is even, and the other is odd. In fact, we have seen this in 
(\ref{eq:trans1}). This immediately implies that the hopping elements
between two sites are only between the functions with equal parity, and we have
two hopping parameters only: $t$ for the "a" orbitals, and $t^{\prime}$ 
for the "b" orbitals
\begin{eqnarray}
{\cal H}_{\rm hop} = -t \sum_{\sigma}c^{\dagger}_{i,a,\sigma}
c_{j,a,\sigma} -t^{\prime}\sum_{\sigma}
c^{\dagger}_{i,b,\sigma}c_{j,b,\sigma} + H.c.
\label{eq:hop1}
\end{eqnarray}
where $\sigma$ is the spin index. In the other directions the hopping 
amplitudes can be obtained by a suitable rotation of the basis functions 
and the hopping matrix. Let us note that for pairs with a different 
orientation, inter-orbital hopping terms will be generated.

``a'' and ``b'' need not mean strictly Ni $d$-states but rather more extended 
one-electron states of the same symmetry. Since one of the main pathways of 
electron propagation would be through the oxygen network, we should think 
of the orbitals as  hybridized Ni-centered Wannier orbitals, but with hopping 
amplitudes which follow not only from Ni--O--Ni hybridization, but by 
considering all finite-amplitude processes which are symmetry-allowed, and 
which in the end-effect may be indexed in the same way as the simple 
nearest-neighbour Ni--Ni hopping. 

It was noted by Mostovoy and Khomskii\cite{most} that the assumption 
of exactly 90$^{\rm o}$ Ni--O--Ni bond angle results in a 
peculiar form of the spin--orbital effective hamiltonian. 
In particular, spin--spin coupling is exclusively ferromagnetic, and orbital 
exchange predominates. One of the ways to look at the situation is that, with 
an ideal octahedron of oxygen atoms, one-electron terms would not allow the 
propagation of an electron from a Ni site to another Ni site via an intervening
oxygen atom. However, other off-diagonal elements, like the spin flip part of
the $p$-shell Hund coupling, still allow electron propagation, and a 
corresponding term in spin--orbital exchange\cite{most}. 
This model may be used to describe NaNiO$_2$, but it is certainly not 
applicable to LiNiO$_2$. Since we aim at deriving both kinds of behavior 
from formally the same hamiltonian, we have to pay particular attention 
to the sources of deviation from the Mostovoy--Khomskii scheme. 

Dar\'e et al.\cite{dare} pointed it out that the trigonal 
splitting of the oxygen $p$ orbitals, and the deviation of the 
Ni--O--Ni bond angle from 90$\,^{\circ}$, facilitate the appearance of 
antiferromagnetic Ni--Ni interactions. However, they did not systematically 
 explore the phase diagram, and neglected several effects which we think 
are important: the direct overlap of the Ni wave functions at 
neighbouring sites, and the intra-atomic exchange and double hopping terms of 
the $d$--$d$ interaction at Ni sites\cite{darenote}. Our aim is a systematic 
investigation of the phase diagram in the entire parameter range.

The form of the on-site pair interaction term ${\cal H}_{\rm Coul}$ is 
restricted by the symmetry classification of the two-electron states:   
${\cal D}_{3d}$ for the orbital component of the wave function, and SU(2) 
for the spin part (which readily gives three singlets and a triplet).   
Orbital quantum numbers follow from $E{\otimes}E = A_1+A_2+E$. 
The anti-symmetrical $A_2$ can be taken with 
symmetrical spin states, yielding the triplet
\begin{eqnarray}
|F_1\rangle & = & c^{\dagger}_{a,\uparrow}c^{\dagger}_{b,\uparrow}|0\rangle
\nonumber\\
|F_2\rangle & = & \frac{1}{\sqrt{2}}(
c^{\dagger}_{a,\uparrow} c^{\dagger}_{b,\downarrow}+
c^{\dagger}_{a,\downarrow} c^{\dagger}_{b,\uparrow})
|0\rangle  
\nonumber\\[1mm]
|F_3\rangle  & = & c^{\dagger}_{a,\downarrow}
c^{\dagger}_{b,\downarrow}|0\rangle\, .
\end{eqnarray}
$A_1$ and $E$ are symmetrical, thus there must be two singlet levels: 
the non-degenerate $A_1$, and the 
two-fold degenerate $E$. The $E$ basis functions are 
\begin{eqnarray}
 |F_4\rangle & = & \frac{1}{\sqrt{2}}(
c^{\dagger}_{a,\uparrow} c^{\dagger}_{b,\downarrow}-
c^{\dagger}_{a,\downarrow} c^{\dagger}_{b,\uparrow})
|0\rangle  
\nonumber\\[2mm]
|F_5 \rangle & = & \frac{1}{\sqrt{2}}(
c^{\dagger}_{a,\uparrow} c^{\dagger}_{a,\downarrow}-
c^{\dagger}_{b,\uparrow} c^{\dagger}_{b,\downarrow})
 |0\rangle\, .
\end{eqnarray}
and the $A_1$ basis function is
\begin{eqnarray}
|F_6 \rangle  =  \frac{1}{\sqrt{2}}(
c^{\dagger}_{a,\uparrow} c^{\dagger}_{a,\downarrow}+
c^{\dagger}_{b,\uparrow} c^{\dagger}_{b,\downarrow})
 |0\rangle\, .
\end{eqnarray}
Their transformation scheme under ${\cal C}_3$
\begin{eqnarray}
 |F_4'\rangle &=& -\frac{1}{2}|F_4\rangle + \frac{\sqrt{3}}{2}|F_5\rangle 
\nonumber\\
 |F_5'\rangle &=& -\frac{\sqrt{3}}{2}|F_4\rangle -\frac{1}{2}|F_5\rangle 
\nonumber
\end{eqnarray}
follows from (\ref{eq:trans2}).

The most general on-site two-body Hamitonian describing the $E{\otimes}E$ 
set of levels is
\begin{eqnarray}
&{\cal H}_{\rm Coul}& = 
  \frac{\tilde U}{2}  n^2 
 - J_H  ({\bf S}_{a} {\bf S}_{b}
              +\frac{3}{4} n_{a} n_{b}) 
\nonumber\\
&& + J_p  ( c^{\dagger}_{a,\uparrow} c^{\dagger}_{a,\downarrow} + 
           c^{\dagger}_{b,\uparrow} c^{\dagger}_{b,\downarrow})
  ( c^{\phantom{\dagger}}_{a,\downarrow} 
c^{\phantom{\dagger}}_{a,\uparrow} + 
      c^{\phantom{\dagger}}_{b,\downarrow}
      c^{\phantom{\dagger}}_{b,\uparrow})
\label{eq:coulham}
\end{eqnarray}
where the ${\tilde U}$ is the familiar on-site repulsion of the Hubbard model,
$J_H$ is the Hund's coupling and $J_p$ is the pair hopping amplitude.
The spectrum of ${\cal H}_{\rm Coul}$ consists of a triplet level at 
${\tilde U} - J_H$ ($|F_1 \rangle$, $|F_2 \rangle$ and $|F_3 \rangle$), 
a twofold degenerate singlet at ${\tilde U}$ 
($|F_4 \rangle$ and $|F_5 \rangle$), and a 
non-degenerate singlet at ${\tilde U}+ 2 J_p$ ($|F_6 \rangle$). 

Since each of the single site terms is invariant under rotations in 
the orbital space, ${\cal H}_{\rm Coul}$ written in (\ref{eq:coulham}) is 
quite general, and its two independent parameters $J_H/{\tilde U}$ and 
$J_p/{\tilde U}$ could be chosen arbitrarily. We may think of these as 
effective interaction parameters which encompass all allowed processes 
affecting the $E$ level under consideration. According to the usual 
evaluation of the simple Coulomb interaction we get $J_p=J_H/2$. 
This physically motivated assumption was used by Castellani et al. 
in their pioneering work\cite{cast} on V$_2$O$_3$. See 
Ref.~\onlinecite{olesrev} 
for further discussions of this point. 

\subsection{The effective hamiltonian from symmetry considerations}

The four-dimensional Hilbert space of $E^1$ states supports 15 local order
parameters \cite{fczhang}. Their standard choice is $S^x$, $S^y$, $S^z$ for
the spins, $T^x$, $T^y$, $T^z$ for the orbitals, and further nine operators 
$S^xT^x$, $S^xT^y$, ... of mixed spin--orbital character\cite{ops}. 
Here we introduced
the $T=1/2$ pseudospin operators
\begin{eqnarray}
  T^x_i &=& \frac{1}{2} \sum_\sigma 
\left( 
  c^{\dagger}_{i,a,\sigma}
  c^{\phantom{\dagger}}_{i,b,\sigma}
  + c^{\dagger}_{i,b,\sigma}
  c^{\phantom{\dagger}}_{i,a,\sigma} 
\right)  
\nonumber\\
  T^y_i &=&  \frac{1}{2 i} \sum_\sigma 
\left( 
  c^{\dagger}_{i,a,\sigma}
  c^{\phantom{\dagger}}_{i,b,\sigma}
  - c^{\dagger}_{i,b,\sigma}
  c^{\phantom{\dagger}}_{i,a,\sigma} 
\right)    
\nonumber\\
  T^z_i &=&   \frac{1}{2} \sum_\sigma 
\left( 
  c^{\dagger}_{i,a,\sigma}
  c^{\phantom{\dagger}}_{i,a,\sigma}
  - c^{\dagger}_{i,b,\sigma}
  c^{\phantom{\dagger}}_{i,b,\sigma} 
\right)    \label{eq:pseudospin}
\end{eqnarray}

For the present, we exploit the separation of spin and orbital Hilbert spaces,
and do not discuss the mixed order parameters, though they are certain to be
as relevant as ${\bf S}$ and ${\bf T}$ in high-symmetry 
situations. 
The symmetry classification of the orbital order parameters is obtained by 
representing the point group ${\cal D}_{3d}$ on the basis of the order
parameters. In fact, since $T^x$, $T^y$, and $T^z$ are composed as 
$c^{\dagger}_{\alpha}c_{\beta}$, the representation we seek is the product 
of the representation (\ref{eq:trans1})--(\ref{eq:trans2}) with its adjoint,
and the decomposition $E{\otimes}E = A_1+A_2+E$ can be used again. It turns
out that $T^x$ and $T^z$ form the basis of the irrep $E$ (a quadrupolar
doublet), while $T^y$ transforms according to $A_2$. We quote the
transformation of the quadrupole operators under the ${\cal C}_3$ rotation 
\begin{eqnarray}
 T'^x &=& -\frac{1}{2} T^x  + \frac{\sqrt{3}}{2} T^z 
\nonumber\\
  T'^z &=& -\frac{\sqrt{3}}{2} T^x -\frac{1}{2} T^z
\label{eq:trans3}
\end{eqnarray}
 From (\ref{eq:trans1}) it is clear that 
\begin{equation}
{\cal C}_2 T^x = -T^x, {\mbox{\ \ \ \ \ }} {\cal C}_2 T^y = -T^y,
{\mbox{\ and \ \ \ \ \ }} {\cal C}_2 T^z = T^z \, .
\label{eq:trans4}
\end{equation}

Finally, let us mention that $T_x$ and $T_z$ are time-reversal 
invariant. The fact that under the time reversal transformation ${\cal T}$, 
${\cal T}T_x=T_x$ and ${\cal T}T_z=T_z$, shows that these are quadrupolar 
order parameters. On the other hand, for the pure imaginary operator $T^y$, 
${\cal T}T_y=-T_y$. In the usual treatment of a cubic $e_g$ doublet, 
$T^y$ would be an octupolar order parameter. However, under trigonal symmetry, 
 $A_2$ is also assigned to the dipolar order parameter $L_{111}$ (orbital 
angular momentum along the 111 direction). Thus our $T^y$ must be a mixed 
dipolar--octupolar order parameter, but we will not analyze its nature in 
detail. 

The form of the effective pair interaction is restricted by the geometrical 
symmetries of the pair, and the nature of the order parameters 
(\ref{eq:pseudospin}). We consider a pair of sites 1 and 2 connected by the 
${\cal C}_2$ axis which figured in our previous considerations. The other 
symmetry element is the perpendicular mirror plane $\sigma_h$ bisecting 
$(1,2)$. 
 
The orbital component of the lowest order effective hamiltonian consists of 
terms $T_1^{\alpha} T_2^{\beta}$ ($\alpha,\beta = x,y,z$), and also of 
single-site terms like $T_1^{\alpha}+T_2^{\alpha}$ (reflecting that the choice 
of the basis is tied to this particular ${\cal C}_2$ axis). The pair energy 
expression must be invariant under ${\cal C}_2$, ${\cal T}$, and also  
$\sigma_h$. $\sigma_h$ acts like
\begin{equation}
\sigma_h T_1^x = -T_2^x, {\mbox{\ \ \ \ \ }} \sigma_h T_1^y = -T_2^y,
{\mbox{\ and \ \ \ \ \ }} \sigma_h T_1^z = T_2^z \, .
\label{eq:trans5}
\end{equation}
 
Time-reversal invariance excludes terms like $T_1^xT_2^y$, and also 
$T_1^y+T_2^y$, and either (\ref{eq:trans4}) or (\ref{eq:trans5}) exclude 
$T_1^x+T_2^x$. In addition, (\ref{eq:trans4}) excludes also $T_1^xT_2^z$. 
Thus we are left with    
\begin{equation}
{\cal H}_{12}^{\prime} = A_x T_1^xT_2^x +  \tilde A_y T_1^yT_2^y 
+ \tilde A_z T_1^zT_2^z + A'_z(T_1^z+T_2^z)\, . 
\label{eq:effham1}
\end{equation}
where $A_x$, $\tilde A_y$, $\tilde A_z$, and $A'_z$ are some real coefficients.
Let us emphasize that, in general, the coupling term $T_1^yT_2^y$ may appear 
in the hamiltonian. 

 Once we introduce spins in the problem, the same arguments hold as 
above with or without spin exchange, so the Hamiltonian becomes
\begin{eqnarray}
{\cal H}_{12} &=& A T_1^xT_2^x +  \tilde A_y T_1^yT_2^y 
+ \tilde A_z T_1^zT_2^z + A'_z(T_1^z+T_2^z)\nonumber\\
 && + \bigl[B' + B T_1^xT_2^x +  \tilde B_y T_1^yT_2^y + 
\tilde B_z T_1^zT_2^z + \nonumber\\
 && + B'_z(T_1^z+T_2^z)\bigr] {\bf S}_1 {\bf S}_2 \, . 
\label{eq:effham1s}
\end{eqnarray}

\begin{figure}[ht]
\begin{center}
\includegraphics*[width=2.5truecm,angle=0]{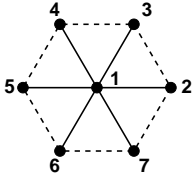}
\caption{The six neighbors of site 1.\label{fig:1234567}}
\end{center}
\end{figure}

Now we consider bonds with different orientation. 
Lattice site 1 has six nearest neighbours 2, 3, 4, 5, 6, and 7, which 
form a regular hexagon (Fig.~\ref{fig:1234567}). The interaction energy 
must be the same for the 
pairs (1,2), (1,3), etc., but just for this reason, the form of the pair 
hamiltonians cannot be. They have to be derived from ${\cal H}_{ij=12}$ by 
suitable transformations.

Consider first the (1,5) pair, which is the mirror image (either through a 
$\sigma_h$ plane containing 1, or by inversion through site 1) of the 
(1,2) pair. 
We can use (\ref{eq:trans5}) to deduce $T_2^x \rightarrow -T_5^x$, 
$T_2^y \rightarrow -T_5^y$, and $T_2^z \rightarrow T_5^z$, thus 
${\cal H}_{ij=15}$ is of the same form as ${\cal H}_{ij=12}$.

The (1,4) pair interaction can be deduced by the ${\cal C}_3$ rotation 
[Eq.~(\ref{eq:trans3})] from Eq.~(\ref{eq:effham1s}). In fact,
\begin{eqnarray}
 T_1^x & \rightarrow & -\frac{1}{2} T_1^x  + \frac{\sqrt{3}}{2} T_1^z 
{\mbox{\ \ \ \ \ }}
T_2^x\rightarrow  -\frac{1}{2} T_4^x  + \frac{\sqrt{3}}{2} T_4^z
\nonumber\\[2mm]
  T_1^z &\rightarrow & -\frac{\sqrt{3}}{2} T_1^x -\frac{1}{2} T_1^z
{\mbox{\ \ \ \ \ }}
T_2^z \rightarrow -\frac{\sqrt{3}}{2} T_4^x -\frac{1}{2} T_4^z
\label{eq:trans6}
\end{eqnarray}
which is more conveniently written as
\begin{equation}
{\bf T}_1 \rightarrow {\bf n}_{14}{\cdot}{\bf T}_1
{\mbox{\ \ \ \ \ }}
{\bf T}_2 \rightarrow {\bf n}_{14}{\cdot}{\bf T}_4
\label{eq:trans7}
\end{equation}
with $
\ {\bf n}_{14}=\left(
\begin{array}{ccc}
-\frac{1}{2} & 0 & \frac{\sqrt{3}}{2}\\
0 & 1 & 0\\
-\frac{\sqrt{3}}{2}& 0 & -\frac{1}{2}
\end{array}\right)
$\\
\\
and the column vector ${\bf T}=(T^x,T^y,T^z)$.\\ 
\\
Similarly, 
$
\ {\bf n}_{15}=\left(
\begin{array}{ccc}
-1 & 0 & 0\\
0 & -1 & 0\\
0 & 0 & 1
\end{array}\right)
$\\

For pairs of different orientation (e.g., (1,3)), analogous expressions can 
be given. For the pair $ij$, the effective hamiltonian ${\cal H}_{ij}$ could 
be deduced from (\ref{eq:effham1s}) by replacing $T^{\alpha}\rightarrow 
({\bf n}_{ij}{\bf T})^{\alpha}={\bf n}_{ij}^\alpha{\bf T}$ everywhere, where
the ${\bf n}_{ij}^\alpha$ with $\alpha=x,y,z$ denotes the first, second or
third row of matrix ${\bf n}_{ij}$, respectively.
However, it is worth to note the following 
simplification. Under these transformations 
$T_1^yT_2^y\rightarrow T_i^yT_j^y$, and $T_1^xT_2^x+T_1^zT_2^z
\rightarrow T_i^xT_j^x+T_i^zT_j^z$. In other words, the orbital base changing 
transformations have the character of a rotation about the pseudospace $y$ 
axis. Pseudospin-space symmetry is easier to identify if in the first line of 
Eq.~(\ref{eq:effham1s}) we make the following rearrangement: 
\begin{eqnarray}
&&A T_i^xT_j^x +\tilde A_y T_i^yT_j^y + \tilde A_z T_i^zT_j^z 
\nonumber \\
&&=
A {\bf T}_i {\bf T}_j + (\tilde A_y-A) T^y_i T^y_j + (\tilde A_z-A) T_i^zT_j^z
\nonumber \\
&&=
A {\bf T}_i {\bf T}_j + A_y T^y_i T^y_j + A_z T_i^zT_j^z \,.
\end{eqnarray}
${\bf T}_i {\bf T}_j$ and $T^y_i T^y_j$ are invariant, so from the 
intersite interaction terms, only the coefficient  of  the $T_i^zT_j^z$ 
term depends on the orientation of the pair. The effective pair 
Hamiltonian is then:
\begin{widetext}
\begin{eqnarray}
\mathcal{H}_{ij} &=& 
  \left[
    A {\bf T}_i {\bf T}_j +A_y T^y_i T^y_j 
    + A_z ({\bf n}_{ij}^z{\bf T}_i) ({\bf n}_{ij}^z{\bf T}_j) 
  + A'_z\left({\bf n}_{ij}^z{\bf T}_i
+ {\bf n}_{ij}^z{\bf T}_j \right) 
    \right] 
\nonumber\\
 && +
  \left[ B'+
  B {\bf T}_i {\bf T}_j + B_y T^y_i T^y_j 
    + B_z ({\bf n}_{ij}^z{\bf T}_i) ({\bf n}_{ij}^z{\bf T}_j) 
  + B'_z\left( {\bf n}_{ij}^z{\bf T}_i
+{\bf n}_{ij}^z{\bf T}_j \right) 
    \right] {\bf S}_i {\bf S}_j \,.
\label{eq:effham3ph}
\end{eqnarray}
\end{widetext}
 
 For completeness, we list here ${\bf n}_{ij}^z$ in all the six possible 
directions:
\begin{eqnarray}
{\bf n}_{12}^z &=& {\bf n}_{15}^z = (0,0,1) \\
{\bf n}_{13}^z &=& {\bf n}_{16}^z = \left(
    \frac{\sqrt{3}}{2}, 0,-\frac{1}{2}
\right) \\
{\bf n}_{14}^z &=& {\bf n}_{17}^z = \left(
    -\frac{\sqrt{3}}{2}, 0,-\frac{1}{2}
\right) 
\end{eqnarray}

The lattice hamiltonian 
\begin{equation}
{\cal H} = \sum_{\langle i, j \rangle} {\cal H}_{ij}
\label{eq:effham4}
\end{equation}
is the sum of (\ref{eq:effham3ph}) over all 
nearest-neighbour pairs.
In the summation the $A'_z$ coefficient drops out as
${\bf n}_{12}^z{\bf T}_1+{\bf n}_{14}^z{\bf T}_1
+{\bf n}_{16}^z{\bf T}_1 =0 $.
The Hamiltonian above contains pure orbital couplings, pure Heisenberg spin 
exchange, and also terms of coupled spin--orbital character.
The only symmetries of the lattice 
hamiltonian are global SU(2) for the spins, and the space group symmetry.  
 
We note here that our theory can be applied to any triangular $d^1$ system 
whose local Hilbert space is the trigonal doublet $E$. The structure of 
BaVS$_3$ can be envisaged as the sequence of triangular 
planes of V$^{4+}=3d^1$ ions. It has been argued that even the minimal model 
of BaVS$_3$ should include the orbital degrees of freedom\cite{bavs}.  
If one assumes that the lowest-lying crystal field level is the $E$ doublet 
derived from the trigonal splitting of $t_{2g}$, our present considerations 
become relevant for BaVS$_3$ as well. 

\vskip.5cm 
\subsection{The effective hamiltonian from microscopic model}

Symmetry considerations do not allow to obtain relationships between the $A$ and $B$ coefficients; they may be derived from the model (\ref{eq:hop1}) 
and (\ref{eq:coulham}) by second-order large-$U$ perturbation theory, as usual for Kugel--Khomskii hamiltonians\cite{KK}. As a result, we get
\begin{widetext}
\begin{eqnarray}
\mathcal{H}_{ij} &=& 
  -\frac{2}{\tilde U\!+\!2J_p} \left[
    2 t t' {\bf T}_i {\bf T}_j
  - 4 t t' T^y_i T^y_j 
    + (t-t')^2 ({\bf n}_{ij}^z{\bf T}_i) ({\bf n}_{ij}^z{\bf T}_j) 
  + \frac{1}{2} (t^2-t'^2) \left( {\bf n}_{ij}^z{\bf T}_i
+ {\bf n}_{ij}^z{\bf T}_j \right) +  \frac{1}{4}(t^2+ t'^2)  
    \right] \mathcal{P}_{ij}^{S=0} 
\nonumber\\
 && -\frac{2}{\tilde U} \left[
    4 t t' T^y_i T^y_j 
  + \frac{1}{2}(t^2+ t'^2)  
  + \frac{1}{2} (t^2-t'^2) \left(
  {\bf n}_{ij}^z{\bf T}_i + {\bf n}_{ij}^z {\bf T}_j 
\right)
\right] \mathcal{P}_{ij}^{S=0} 
\nonumber\\
 && -\frac{2}{\tilde U\!-\!J_H} \left[  - 2 t t' {\bf T}_i {\bf T}_j
   - 
   (t-t')^2 ({\bf n}_{ij}^z{\bf T}_i) ({\bf n}_{ij}^z{\bf T}_j)+
\frac{1}{4}(t^2+t'^2)
    \right] \mathcal{P}_{ij}^{S=1} 
\label{eq:effham3}
\end{eqnarray}
\end{widetext}
We found it convenient to express the Hamiltonian using the
 $\mathcal{P}_{ij}^{S=0}$ and $\mathcal{P}_{ij}^{S=1}$ projection
operators onto the singlet and triplet spin combination on the bond:
\begin{equation}
\mathcal{P}_{ij}^{S=0} = \frac{1}{4} - {\bf S}_i{\bf S}_j 
\quad\mbox{and}\quad
\mathcal{P}_{ij}^{S=1} = {\bf S}_i{\bf S}_j+\frac{3}{4}
\end{equation}
First, some general remarks about the parameter range. (\ref{eq:coulham}) 
shows a two-parameter manifold of on-site Coulomb hamiltonians. However, 
we do not change $J_p/J_H$ continuously, but investigate two special cases 
only: (a) neglecting pair hopping $J_p=0$ (a frequent, though not clearly 
motivated, simplification), and (b) the physically motivated choice 
$J_p=J_H/2$. Most of our results will be about the latter case, using 
the notation $J=2J_p=J_H$.

Redefining the basis states $\phi_a\leftrightarrow\phi_b$ interchanges the 
definitions of $t$ and $t^{\prime}$, thus it is sufficient to consider the 
$|t|>|t'|$ case.
 
It is, however, worth noting that the orbital part of (\ref{eq:effham3}) 
becomes SU(2) invariant for $t=t'$ and $J_p=0$:
\begin{eqnarray}
\mathcal{H}_{ij} &=& 
  \frac{4t^2}{\tilde U} 
  \left( {\bf T}_i {\bf T}_j + \frac{3}{4} \right) 
  \left({\bf S}_i{\bf S}_j -  \frac{1}{4}\right) 
\nonumber\\
 &&+ \frac{4 t^2}{\tilde U\!-\!J_H}
   \left({\bf T}_i {\bf T}_j -  \frac{1}{4}\right) 
   \left({\bf S}_i{\bf S}_j+\frac{3}{4} \right)\, .
\label{eq:pairsu2}
\end{eqnarray}
The lattice hamiltonian has now global SU(2) symmetry for the spins {\sl and} 
global SU(2) symmetry for the pseudospins (global SU(2)$\otimes$SU(2), with 
the six conserved quantities $\sum_jS_j^{\alpha}$, $\sum_jT_j^{\beta}$, 
for $\alpha,\beta=x,y,z$).

A still higher symmetry is obtained for $J_H=J_p=0$ 
when the pair Hamiltonian simplifies to the SU(4) symmetrical\cite{fczhang} 
\begin{equation}
\mathcal{H}_{ij} =
  \frac{8t^2}{\tilde U} 
  \left( {\bf T}_i {\bf T}_j + \frac{1}{4} \right) 
  \left({\bf S}_i{\bf S}_j + \frac{1}{4}\right)\, . 
\label{eq:pairsu4}\end{equation}
The corresponding lattice hamiltonian possesses global SU(4) symmetry (there are fifteen 
conserved quantities: $\sum_jS_j^{\alpha}$, $\sum_jT_j^{\beta}$, and  
$\sum_jS_j^{\alpha}T_j^{\beta}$ for $\alpha,\beta=x,y,z$). 

\section{Ground states of the pair and the tetrahedron problem}

In what follows, we seek to find the possible different types of ground 
state of (\ref{eq:effham4}) on the triangular lattice. For a first 
orientation, we describe the results for small systems, then go over to 
larger ones. Whenever possible, we use preconception-free numerical methods, 
and then try to re-interpret the results with approximate theories which can, 
in principle, be generalized to infinite system size. It is a general 
trend that with increasing system size, complicated states are found whose 
existence could not have been guessed by simple-minded extrapolation from 
small systems. Therefore we will have to be cautious in drawing conclusions 
about the thermodynamic limit.

Before we turn to the physically motivated  $J_p=J_H/2$ case, we examine the case 
when the pair hopping amplitude is absent.

\subsection{Two site problem}

For simple spin models, the correlations found for a pair of sites allow to 
infer the character of the ordered phase in the thermodynamic limit\cite{afm}. 
Our first aim is to map the pair solutions, and try to deduce how spin and 
orbital order may complement each other.

The most notable consequence of setting $J_p=0$ is that the 
$4tt^{\prime}T_i^y T_j^y$ term cancels from the first and second row of  
 the effective Hamiltonian (\ref{eq:effham3}). Naturally, there is still a 
$T_i^y T_j^y$ interaction included in the isotropic term 
${\bf T}_i {\bf T}_j$. On this basis, one may not expect a preference for 
$T^y$-polarized (complex) orbital ground states. However, one should not 
overlook the possibility that the system may choose $T^y$-polarization 
as a compromise when interaction terms preferring real orbital order 
mutually frustrate each other~\cite{typol}.  
 
Let us note that the hamiltonian of the $ij=12$ bond
\begin{widetext}
\begin{eqnarray}
{\cal H}_{12} & = & -\frac{2}{\tilde U} \left[ 2tt^{\prime}\left( 
T_1^xT_2^x + T_1^yT_2^y \right) + (t^2+ t^{{\prime}2}) T_1^zT_2^z + 
(t^2 - t^{{\prime}2}) (T_1^z+T_2^z) +\frac{3}{4}(t^2 + t^{{\prime}2})\right]
{\cal P}_{12}^{S=0}\nonumber\\[2mm]
& & -\frac{2}{{\tilde U} - J_H} \left[-(t^2 + t^{{\prime}2})T_1^zT_2^z 
   - 2tt^{\prime}\left( T_1^xT_2^x + T_1^yT_2^y \right)
+\frac{1}{4} (t^2 + t^{{\prime}2})\right]{\cal P}_{12}^{S=1}
\label{eq:effhampair}
\end{eqnarray}
\end{widetext}
has two new symmetries characteristic of the two-site problem. One of them 
is axial symmetry about $T^z$ in pseudospin space\cite{others}, 
which allows to classify 
the eigenstates as $T_1^z+T_2^z$ eigenstates. The other is a the 
$t^{\prime} \leftrightarrow -t^{\prime}$ symmetry: a $\pi$-rotation 
about $T^z$ in pseudospin space for site 2 is a canonical transformation 
which leaves the energy unchanged, but it amounts to 
$t^{\prime} \rightarrow -t^{\prime}$. This symmetry can be restated for 
larger clusters with bipartite structure, but it cannot be extended to 
$N>2$ clusters of the triangular lattice. 

For $t=t^{\prime}$ SU(2)${\otimes}$SU(2) symmetry follows as in 
(\ref{eq:pairsu2}). Taken in conjunction with the previous remarks, 
 the $t=-t^{\prime}$ model must have the same symmetry. Similarly, the 
degeneracies must be the same for the SU(4) point $t=t^{\prime}$, $J_H=0$, 
and its mirror image $t=-t^{\prime}$, $J_H=0$.

The Hilbert space of two electrons on two sites is 16 dimensional, and the
energies and orbital eigenstates for the $ij=12$ bond are:
\begin{equation}
\begin{array}{ccc}
  E_{S=0}=-\frac{4 t^2}{\tilde U}, & E_{S=1}=0, & |aa\rangle \\
  E_{S=0}=-\frac{4 t'^2}{\tilde U}, & E_{S=1}=0, & |bb\rangle \\
  E_{S=0}=-\frac{(t+t')^2}{\tilde U}, 
& E_{S=1}=-\frac{(t-t')^2}{\tilde U-J_H}, & 
  |ab\rangle + |ba\rangle \\
  E_{S=0}=-\frac{(t-t')^2}{\tilde U}, 
& E_{S=1}=-\frac{(t+t')^2}{\tilde U-J_H}, & 
  |ab\rangle - |ba\rangle
\end{array}
\end{equation}

\begin{figure}[ht]
\begin{center}
\includegraphics*[width=7.0truecm]{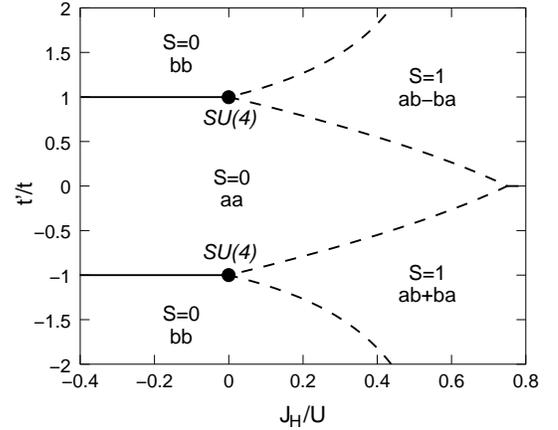}
\caption[diag4K]{\label{fig:diag2K}{Zero temperature phase diagram 
of the model on two sites
    as a function of $t'/t$ and $J_H/\tilde U$. We have indicated the spin and
    orbital parts. Along the thick line at $t'/t=1$ and $J_H<0$ the ground
    state is triply degenerate: the orbital part becomes SU(2) symmetric, and
    the orbital triplet $|aa\rangle$, $|ab\rangle +|ba\rangle $ and
    $|bb\rangle$ with the spin singlet forms the ground state wave function.
    Along the $t'/t=-1$ and $J_H<0$ line, the orbital triplet consists of 
   $|aa\rangle$, $|ab\rangle -|ba\rangle $ and
    $|bb\rangle$. At the SU(4) points the ground state is  6-fold degenerate. 
\label{fig:pair}}}
\end{center}
\end{figure}

The results are shown in Fig.~\ref{fig:pair}. Inside any of the ground state 
phases, either the spins are parallel, and the orbitals antiparallel, 
or vice versa. At the boundaries, the ground state level has higher 
degeneracy which can be interpreted as the manifestation of 
one of the higher symmetries discussed above.

For the spin singlets, $t\ne t'$ acts like an external orbital field, and 
therefore orbital polarization is either in the $a$, or the $b$ direction. At 
the line $t=t'$ (section $J_H<0$) not only $aa$ and $bb$ (states from the 
neighbouring domains) become degenerate but also $|ab\rangle+|ba\rangle$, 
thus the ground 
state is the threefold degenerate spin-singlet-orbital-triplet ($S=0$, $T=1$), 
allowed by the symmetry SU(2)${\otimes}$SU(2). Continuing the $t=t'$ 
line\cite{notethat} to 
$J_H>0$, the ground state is again threefold degenerate, but its nature 
changed to spin-triplet-orbital-singlet ($|ab\rangle-|ba\rangle$). 
The border point 
$t=t'$, $J_H=0$ is the SU(4) point where the ($S=0$, $T=1$), and 
($S=1$, $T=0$) levels become degenerate [forming the basis of the 
six-dimensional anti-symmetrical irrep of SU(4)].

Analogous results hold for the other SU(2)${\otimes}$SU(2) line $t=-t'$, only 
for $J_H>0$, the $|ab\rangle-|ba\rangle$ orbital state ($T=0$, $T^z=0$) is changed to $|ab\rangle+|ba\rangle$ 
($T=1$, $T^z=0$). Note that the energy difference between these two states 
comes from the  term of 
(\ref{eq:effhampair}) proportional to $t'(T_1^+T_2^-+T_1^-T_2^+)$, 
so changing the sign of $t'$ changes the parity of 
the ground state. The sixfold degeneracy of the ground state at the 
``anti-SU(4)'' point $t=-t'$, $J_H=0$ follows from the 
$t' \leftrightarrow -t'$ symmetry of the pair problem.

A very similar phase diagram would be obtained for $J_p\ne 0$. We do not show 
it here, but we include the contribution of pair hopping in all our 
subsequent calculations.

If any far-reaching conclusions from Fig.~\ref{fig:pair} could be drawn, it 
would be that the ground state has either ferro-orbital order and spin 
antiferromagnetism (or a singlet spin liquid); or it is a spin ferromagnet 
with staggered orbital order (or orbital liquid). Less obviously, at the SU(4) 
points, a spin--orbital quantum liquid may be inferred\cite{su4penc}. 

To either confirm, or disprove, these guesses, two routes can be followed: 
a) determine the exact phase diagram of larger clusters and see if there is 
a clear trend emerging; b) construct variational wave functions which possess 
the envisaged correlations. Fig.~\ref{fig:pair} suggests that 
antiferromagnetic effective spin models can be derived easily because 
uniform orbital order factorizes site-by-site. However, for high-spin states, 
the orbital states  are more complicated.
SU(4)-like states, for which spins and orbitals are entangled, pose further 
challenge.

\subsection{Four site problem}

In what follows, we set $J=2J_p=J_H$. 
The four site cluster with periodic boundary conditions is equivalent to a
 tetrahedron where the three directions on the triangular lattice
correspond to the three pairs of 
opposite bonds
 on the tetrahedron. This cluster proves to be sufficiently large to provide 
us with some insight into the problem.

As a first step, we do exact (numerical) diagonalization for the 
hamiltonian
\begin{equation}
{\cal H}_{\rm tetr} = {\cal H}_{12} + {\cal H}_{13} 
+{\cal H}_{14} +{\cal H}_{23} 
+{\cal H}_{24} +{\cal H}_{34}
\label{eq:tetr1}
\end{equation}
in the $4^4=256$-dimensional Hilbert space. The total 
spin
$S$ and its $z$-component 
$S^z$ are good quantum numbers, but this is only of limited use in 
identifying eigenstates. The hamiltonian couples 
spin correlations with orbital correlations, therefore most of the eigenstates 
have mixed spin--orbital character. More precisely: the $S=2$ eigenstates can 
be sought in the factorized form
\[
|\uparrow\uparrow\uparrow\uparrow\rangle{\otimes}\Phi(T_1,T_2,
T_3,T_4)
\] 
but this is no longer true of lower-spin states. In particular, we know that 
there are only two independent spin singlet states, [12][34] and [23][41],
but combined with the orbitals, we have 32 independent $S=0$ 
spin--orbital states. Most of the $S=0$ eigenstates of Eq.~(\ref{eq:tetr1}) 
are 
not represented as a linear combination of the above singlets multiplied by 
a pure orbital state. In fact, an overall singlet which plays a prominent 
role in our considerations is the SU(4) plaquette singlet
\begin{equation}
\Psi_{\rm SU(4)}=[12]\{ 23\}[34]\{41\} - [23]\{ 34\}[41]\{12\}
\label{eq:su4}
\end{equation}
where $\{ 23\}$ represents the pseudospin singlet connecting sites 2 and 
3, etc. 
It is clear that $\Psi_{\rm SU(4)}$ is a spin singlet, just as it is a 
pseudospin singlet, and it does not factorize in spin and orbital variables. 
$\Psi_{\rm SU(4)}$ is the ground state of ${\cal H}_{\rm tetr}$ in 
the SU(4)-symmetrical point ($t=t'$, $J=0$) of the parameter space 
of (\ref{eq:tetr1}). It is the only 
SU(4) singlet (the only basis function for the 1-dim irrep of SU(4)) in the 
present Hilbert space.

\begin{figure}[ht]
\begin{center}
\includegraphics*[width=7.5truecm]{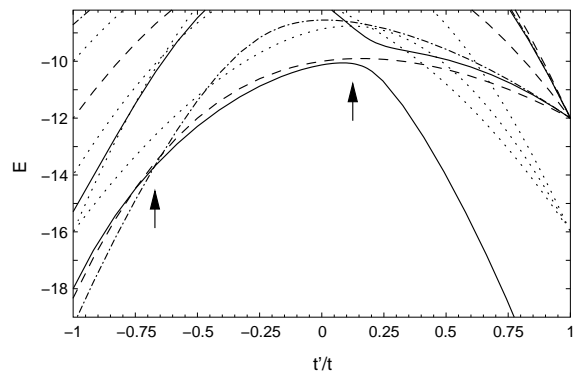}
\caption{\label{fig:enediag4}
Energy spectrum of the spin singlet sector in the tetrahedron as a function 
of $t'/t$ for the $J_H=J_p=0$ case. For $t'/t$ between 0.2 and 1 the ground 
state is well separated from the rest of the states and it is the adiabatic 
continuation of the SU(4) singlet state at the $t'/t=1$ point (full line). 
The SU(4) singlet nature of the ground state is lost at around  $t'/t=0.2$ 
(denoted by an arrow), and in the  
region $-0.7 < t'/t < 0.2$ three levels 
(a non-degenerate level (solid line)  and a 2-fold degenerate level 
(dashed line)) 
go together. Finally, at $t'/t=-0.7$ the symmetry of the ground state changes, 
indicating the appearance of the third phase (dashed-dotted line). }
\end{center}
\end{figure}

We diagonalized (\ref{eq:tetr1}), and followed the low-lying states (a 
representative example is shown in Fig.~\ref{fig:enediag4}). While 
the detailed nature of the ground state always has some continuous dependence 
on the hamiltonian parameters $t'/t$ and $J/U$, there
are also sharp changes at level crossings. A level crossing is possible between 
states with different symmetry labels. The symmetry of ${\cal H}_{\rm tetr}$ 
is SU(2)${\otimes}T_d$, where $T_d$ is the tetrahedral group. $T_d$ has 
one- and three-dimensional irreps; furthermore, granting time-reversal 
invariance, two complex conjugate one-dim irreps belong to degenerate energy 
levels. We distinguish between ground states 
according to their spin degeneracy ($2S+1$), and orbital degeneracy which can 
be 1, 2, or 3. 

\begin{figure}[ht]
\begin{center}
\includegraphics*[width=7.0truecm]{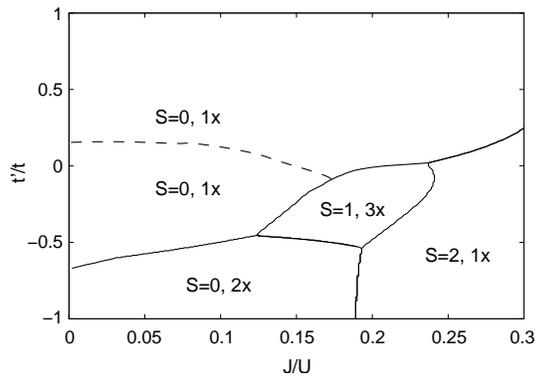}
\caption{\label{fig:diag4}
Phase diagram of the spin--orbital model with $J=J_H=2J_p$ and $U=\tilde U-J$ 
on a tetrahedron, based on exact diagonalization (see also 
Fig.~\protect\ref{fig:enediag4}). Phase boundaries in bold lines belong 
to level crossings in the ground state energy. A further singlet-to-singlet 
transition is identified in the vicinity of an antilevel crossing 
(dashed line). The degeneracy (apart from the trivial spin degeneracy) 
of each state is also indicated. }
\end{center}
\end{figure}

The resulting phase diagram is shown in Fig.~\ref{fig:diag4}. Phase 
boundaries were drawn where we found a clear change in the character of 
the ground state; this holds also for the boundary between the two 
non-degenerate ($S=0$, 1x) phases. Let us immediately point out that the 
tetrahedral phase diagram is very different, and therefore would have 
been difficult to guess, from the pair phase diagram shown in 
Fig.~\ref{fig:pair}. Taken in itself, the lack of mirror symmetry
about the $t'=0$ axis was to be expected, since the tetrahedral cluster is not 
bi-partite. In particular, the $t'=-t$, $J_H=0$ point does not have any 
special significance.
 However, less obvious features are the variety of singlet phases, 
the shrinking of the domain of spin-polarized solutions, and the predominance 
of the singlet phase into which the SU(4) point is embedded.

\section{Variational approach for the four site cluster}

\begin{figure}[ht]
\begin{center}
\includegraphics*[width=7.5truecm]{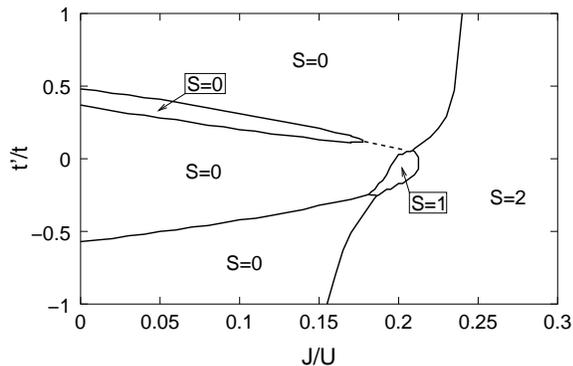}
\caption{\label{fig:MFtetra}
The phase diagram of the effective model on a tetrahedron based on the
spin-orbital decoupling scheme (\protect\ref{eq:STfact}).}
\end{center}
\end{figure}

\subsection{The method}

The previous section showed us that we could expect a rich phase diagram 
for our model even on a small size cluster. We will continue our investigation
by studying larger clusters by a kind of variational method:
since there is a strong asymmetry between the spin and the orbital parts in 
the Hamiltonian, we try to decouple spin and orbital degrees of 
freedom\cite{frederic} 
by factorizing the wave function into 
a $|\Psi^{S}\rangle$ spin 
and $|\Psi^{T}\rangle$ orbital part:
\begin{equation}
|\Psi_{\rm ST}\rangle=|\Psi^{S}\rangle \otimes |\Psi^{T}\rangle\, .
 \label{eq:STfact}
\end{equation}
While this factorization applies to the pair problem, it cannot 
describe the entanglement of spin and orbital fluctuations for $N\ge 4$ 
sites. In particular, it does not allow to capture the SU(4) character 
displayed  by (\ref{eq:su4}). However, it should work well for states with 
weakly fluctuating orbital order. 

We proceed as follows: in the effective Hamiltonian we can separate a 
spin--orbital mixing term, and purely orbital terms:  
\begin{equation}
 \label{h-form}
 H=\sum_{i,j}\left\{2(\bf{S}_i.\bf{S}_j)h^{T}_{ij}+k_{ij}^T\right\}
\end{equation}
Next, we need to minimize the Hamiltonian by using the factorized wave 
function: $ \langle \Psi_{\rm ST} | H |\Psi_{\rm ST}\rangle $. It implies that
$|\Psi_{\rm S}\rangle$ is an eigenstate of the Hamiltonian
\begin{equation}
 \sum_{i,j} 2(\bf{S}_i.\bf{S}_j) \langle \Psi_{\rm T} | h^{T}_{ij}|\Psi_{\rm T}\rangle 
\end{equation}
while 
$|\Psi_{\rm T}\rangle$ is an eigenstate of the Hamiltonian
\begin{equation}
 \sum_{i,j}\left( 2 \langle \Psi_{\rm S}| {\bf S}_i {\bf S}_j |\Psi_{\rm S} \rangle  h^{T}_{ij} + k^{T}_{ij}\right) 
\end{equation}
This coupled set of equations is solved by iteration, 
keeping at each step the solutions with lowest eigenvalue.

We have applied this technique to obtain the phase diagram of the regular
 4-site (tetrahedron) and 16-site cluster.
 
\subsection{The phase diagram on a tetrahedron}

As we can see in Fig.~\ref{fig:MFtetra}, the "mean-field" phase diagram 
shows a remarkable resemblance to the exact one (Fig.~\ref{fig:diag4}). 
We should, however, note that 
the ferromagnetic region extended too much at the expense of the SU(4) 
phase, basically because the Ansatz (\ref{eq:STfact}) cannot describe SU(4) 
correlations\cite{su4but}, 
while the spin-aligned states are treated correctly.  
 The $S=1$ region has shrunk, too. Our variational recipe for $S=1$ states 
is to compose them of two bonds: a spin triplet and orbital 
$|ab\rangle + |ba\rangle$ bond, and a spin singlet and 
orbital $|aa\rangle$ type bond. These can be permuted and rotated to give 6
solutions which are degenerate at the mean-field level. Allowing for the
resonance between these six states, we can reproduce the 3-fold 
degenerate  $S=1$ state seen in exact-digonalization study by taking the 
appropriate linear combinations of them.

 In the singlet sector we can distinguish between several phases: The lowest
 one is composed of spin triplet and orbital $|ab\rangle+|ba\rangle$-like
 bonds, which are composed into a singlet, and is 3-fold degenerate at the
 mean-field level due to
 possible rotations (here again, the off diagonal matrix elements between the
 states will favor the 2-fold degenerate linear combination, in agreement with
 Fig.~\ref{fig:diag4}). 
In the remaining part, the spin wave functions is the 
same (singlet valence bonds), only the orbital character changes from
 $|aa\rangle$ type bonds to a more complicated one close to the $t=t'$, 
$J=0$ SU(4)-symmetric point (where the approach we use is clearly not 
applicable). The
 number of solutions of the iteration becomes very large in the vicinity of 
the SU(4) point, with essentially the same energy. 

\section{Variational approach for the 16-site cluster}

The tetrahedron solutions show that there must be quite a few phases with 
markedly different spin--orbital correlations. However, the $N=4$ cluster 
is too small to draw inferences about the character of any emerging long 
range order (except for spin ferromagnetism). Therefore we investigated an 
$N=16$ cluster which is large enough to differentiate between 
quasi-one-dimensional (chains) and genuinely two-dimensional orbital ordering 
patterns. We use the same variational method as for $N=4$.   

\begin{figure}[ht]
\begin{center}
\includegraphics*[width=7truecm,angle=0]{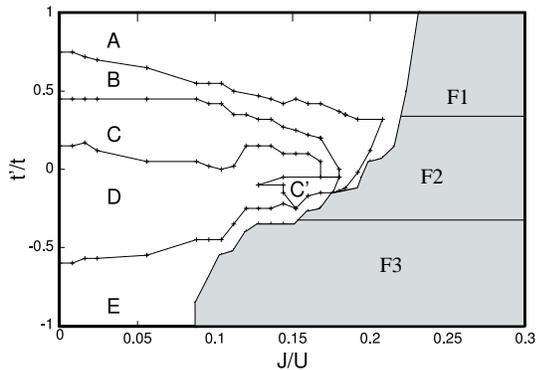}
\caption[var-dia]{\label{var-dia}{Mean-field phase diagram on a 16-site cluster
as a function of hopping 
integral versus Hund's coupling. The grey phase is the ferromagnetic phase,
with the classical phase boundaries shown (see section \ref{Vb})}}
\end{center}
\end{figure}

 As shown in 
Fig.~\ref{var-dia}, the model leads to a rich phase diagram. For reasonably 
large values of $J/U$ we find the fully polarized ferromagnetic ($S=8$) 
region with three phases that differ by their orbital structure. 
In the spin singlet region we can again 
distinguish at least 6 phases, which are labeled by capital letters. 
A detailed 
discussion of these phases follows.

\begin{figure}[ht]
\begin{center}
\includegraphics*[width=5truecm,angle=0]{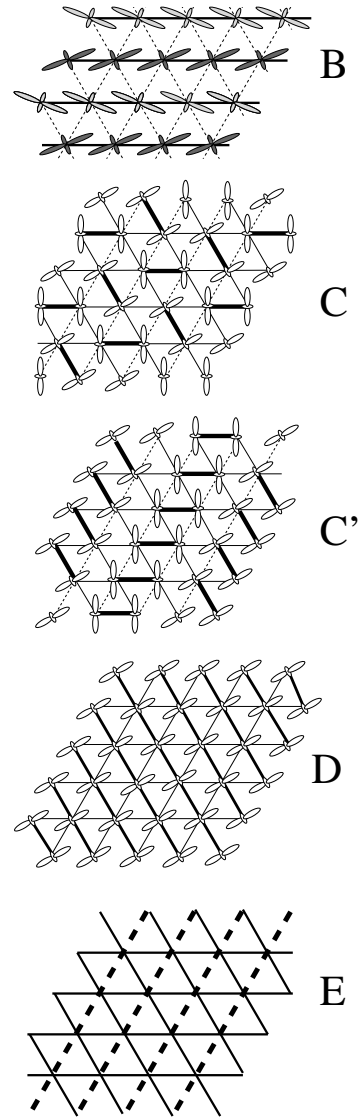}
\caption[var-dia]{\label{phases}{Spin and orbital structure in the singlet
phases of the mean-field phase diagram. Solid line indicates AF, dashed line
FM spin correlations.}}
\end{center}
\end{figure}

\subsection{The singlet phases}

The $S=0$ part of the diagram is composed of 6 phases. We have investigated 
in more details each phase starting from $t'/t\sim -1$ and going through 
the 4 boundaries until $t'/t\sim 1$ for several ratios $J/U$.
The aim of this section is to understand the different types of orbital and 
spin orders. The nature of the spin phases turned out to be easily determined 
from the variational method itself: In all cases except phase A,
some clear pattern
with large and positive or negative values of 
$\langle \bf{S}_i .\bf{S}_j \rangle$ could be identified, leading to 
magnetic or singlet dimer order. The orbital part was more tricky to
identify since the most relevant operator is not 
$\bf{T}_i.\bf{T}_j$ but $h_{ij}^T$,
and a given mean-value of this parameter does not obviously lead
to an orbital state since this operator is quite involved.
So to get a simple physical picture of the orbital structure we 
have tried in each case to reproduce the pattern given
by the mean-field solution for $\langle h_{ij}^T\rangle$
assuming at each site an orbital wave-function of
the form
\begin{equation}
|\Psi^T_i\rangle=(\cos\theta_i|a_i\rangle+\sin\theta_i|b_i\rangle)
\end{equation}
and we have checked that this orbital structure also reproduces
satisfactorily the mean value of $\bf{T}_i.\bf{T}_j$
measured in the mean-field ground state.
This turned out to give a clear picture in all phases except A and E.
The information obtained in this way is summarized in 
Fig.~\ref{phases}. In the following, we describe in more details
all these phases.

\subsubsection{Phase A}

This phase contains the SU(4) point ($t'=t,J=0$) for which the 
mean-field decoupling used here is known to be inadequate given
the very symmetric roles played by the spin and orbital degrees
of freedom~\cite{frischmuth}. In fact, it is believed that
at the SU(4) point the system is in a spin and orbital 
liquid state involving resonances between SU(4) singlet
plaquettes. A discussion of the physical properties at the
SU(4) point can be found in Ref. [\onlinecite{su4penc}].
Although the mean-field solution is not directly relevant
for that phase, the mean-field approach is still useful
to determine the boundary of the SU(4) region since it 
allows to detect the domain of stability of the neighbouring
phases, for which the mean-field solution is indeed relevant,
as will be discussed below.
As anticipated, the SU(4) physics extends to 
a finite and relatively large portion of the phase diagram,
and it can in principle be relevant for real systems. 
Since our mean-field approach does not lead to any 
physical insight beyond the determination of the boundary
of this phase however, we will not discuss it further here.

\subsubsection{Phase B}

 From the magnetic point of view, this phase consists essentially
of weakly coupled, antiferromagnetic chains (see Fig.~\ref{phases}),
while the orbital structure turns out to be rather subtle with an 
antiferro-orbital arrangement of ferro-orbital chains
with orbitals which are neither pure 
$|a \rangle =|d_{3z^2-r^2}\rangle$ nor
$|b\rangle =|d_{x^2-y^2}\rangle$ but alternate between 
$\frac{1}{\sqrt{2}}(|a\rangle+|b\rangle)$ and 
$\frac{1}{\sqrt{2}} (-|a\rangle+|b\rangle)$.
The detailed magnetic structure depends {\it a priori}
on the residual couplings between the chains. 
If the couplings are equal in both residual directions,
some canting will presumably develop inside the chains
to accomodate the frustration,
like in the limiting case of the 180 degree classical ground
state of the Heisenberg model on the triangular lattice.
This effective magnetic hamiltonian would be similar to that realized
in Cs$_2$CuCl$_4$, with possibly spinon excitations as reported
by Coldea et al\cite{coldea}. If however the symmetry 
is broken between the residual directions, the system is
expected to develop rather collinear order, with
lines of parallel
spins along the direction of the most ferromagnetic or least 
antiferromagnetic residual coupling. 
For all parameters, the residual couplings predicted by
the mean-field solution are very small, but their sign and
symmetry depends on the parameters. They tend to be
AF for small $J$ and ferromagnetic for large $J$, and the 
symmetry between the two directions may or may not
be broken depending on the parameters. While this interesting 
point would deserve further investigation, we do not think that a reliable
answer to such a subtle issue can be obtained just on the basis 
of this mean-field decoupling, and we do not discuss the point
further.

\subsubsection{Phases C and C'}

Both phases are characterized by strong dimer singlets 
forming different regular dimer coverings of the triangular 
lattice. On each dimer the orbitals are parallel, and
they correspond to $d_{3z^2-r^2}$, $d_{3x^2-r^2}$ or 
$d_{3y^2-r^2}$ depending on the orientation of the bond.
Note that all these orbitals are Jahn-Teller active,
leading in all cases to two long bonds and four 
short bonds.
One might be tempted to conclude that these phases
correspond to two types of valence bond solids with 
the patterns depicted in Fig.~\ref{phases}. The mean-field
approach has a very remarkable property however: In 
addition to the mean-field solutions with lowest energy shown
in Fig.~\ref{phases}, there are several other
mean-field solutions of the self-consistent equations
with energies very close to the lowest energy corresponding
to other dimer coverings of the triangular lattice.
In such circumstances, going beyond mean-field is likely
to couple these solutions, and the relevant model would
then be a quantum dimer model describing resonances between
these states. As we shall see below, this point of view
is favoured by exact diagonalizations of finite clusters.
So at that stage we think it is safer to think of these
phases as a region of parameters where all dimer coverings
are relevant states for low-energy physics.

\subsubsection{Phase D}

This phase consists essentially 
of weakly coupled antiferromagnetic
chains, but in contrast to Phase B, the orbital structure is now 
ferro-orbital with
only orbital $d_{3z^2-r^2}$, $d_{3x^2-r^2}$ or 
$d_{3y^2-r^2}$ depending on the overall direction of the AF chains.
Since these orbitals are Jahn-Teller active, one expects
in this case that the system would undergo a cooperative
Jahn-Teller distortion with two long bonds per octahedra
all pointing in the same direction. Like in Phase B, the
actual magnetic structure will be controlled by the residual 
couplings, and all the discussion of Phase B applies here,
including the sign of the residual couplings and the symmetry
of the couplings in the directions of weak coupling. In that
case too, a reliable determination of the possible magnetic
phases requires further investigation that goes beyond
the present mean-field calculation.

\subsubsection{Phase E}

This phase is dominated by strong
antiferromagnetic correlations in two directions and weak
ferromagnetic correlations in the third direction, 
leading to an effective N\'eel structure. The orbital 
structure cannot be reproduced satisfactorily with
the variational ansatz of one orbital wavefunction per site.
The pattern of $\langle h_{ij}^T\rangle$ would be
consistent with a ferro-orbital ordering with orbitals
$\frac{1}{\sqrt{2}}(|a\rangle+|b\rangle)$ at all sites,
but the $\bf{T}_i.\bf{T}_j$ correlations are not
ferromagnetic. So to decide on a possible orbital order
would require to go beyond the present mean-field approach.

\subsection{The ferromagnetic phase}\label{Vb}

In the ferromagnetic region the $\mathcal{P}_{ij}^{S=0}$ spin singlet 
projection is 0, 
so that the effective Hamiltonian (\ref{eq:effham3}) is reduced to the 
following form (neglecting the constant term):
\begin{equation}\label{ferroh}
H_{\rm eff}^{\rm FM}=
 \frac{2}{\tilde U\!-\!J_H}
\sum_{i,j} \left[
   (t-t')^2 ({\bf n}_{ij}^z{\bf T}_i) ({\bf n}_{ij}^z{\bf T}_j)
 +    2 t t' {\bf T}_i {\bf T}_j 
\right]\end{equation}
and the orbital structure only depends on the ratio $t'/t$.
As shown in the phase diagram~\ref{var-dia} we can distinguish three phases 
going from $t'/t=-1$ to $t'/t=1$. All the identified phases identified are 
orbitally ordered
phases. They can be understood starting from the classical limit, which in
our case is equivalent to minimizing the energy of the 
\begin{equation}
|\Psi\rangle=\prod_j 
\left(\cos\theta_j |a\rangle+e^{i\phi_j}\sin\theta_j |b\rangle \right)
\label{eq:orbMFwf}
\end{equation}
site-factorized wave function.  The phase boundaries shown in
  Fig.~\ref{var-dia} are obtained by equating classical energies 
obtained from the wave function of Eq. \ref{eq:orbMFwf}.

\begin{figure}[ht]
\begin{center}
\includegraphics[width=5truecm]{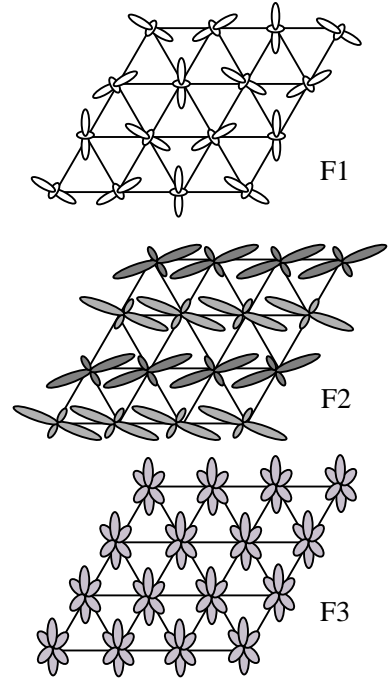}
\caption[tf]{\label{fig:tabfm}{Schematic representation of the orbital
    orderings in the spin ferromagnetic case.}}
\end{center}
\end{figure}

\subsubsection{Phase F1}
For  $t'=t$ the Hamiltonian of the orbitals becomes the  
standard SU(2) symmetric Heisenberg Hamiltonian with antiferromagnetic 
exchange. In this case a three sublattice long-range order for the 
${\bf T}$ pseudospins develops. Away from the SU(2) symmetric point, the 
three sublattice LRO is stable up to $t'=t/3$, with the 120$\,^{\circ}$
configuration  restricted in the $(T^x,T^z)$ plane 
[in Eq.~(\ref{eq:orbMFwf}) we choose $\theta$ for $\theta_j$'s 
in the first, $\theta+2\pi/3$ for $\theta_j$'s in the second, and 
$\theta-2\pi/3$ for $\theta_j$'s in the third sublattice, with
$\phi_j=0$ everywhere] with energy
\begin{equation}
  \frac{E_{\rm AFO}}{N} = -\frac{3}{8}\frac{(t+t')^2}{\tilde U - J_H} \,.
\end{equation}
 We have shown a possible
120$\,^{\circ}$ orbital pattern with $\theta=0$ in the top of the 
Fig.~\ref{fig:tabfm}. While the classical approach does not allow us to fix the
value of $\theta$, this degeneracy is probably lifted by quantum
fluctuations. 

For finite systems, 
the signature of the developing LRO can be found in the energy spectrum in the 
form of the Anderson's tower, as has been confirmed by 
Bernu {\it et al.} for the isotropic triangular lattice \cite{bernu}. 
These low lying states ($\Gamma1$, $K1$ and $\Gamma4$) 
can also be seen in Fig.~\ref{fig:FM_ED}, and they can be continuously
followed up to the isotropic point $t'=t$, where they become the lowest lying 
pseudospin triplet excitations. Further evidence comes from the 
nearest- and  next-nearest neighbour $\langle {\bf T}_i {\bf T}_j \rangle$ 
correlations. There is a strong ferro-orbital correlation between 
a site and its second nearest-neighbours, e.g.   
$\langle {\bf T}_i {\bf T}_j \rangle\sim 0.19$ for $t'/t=0.8$.

\subsubsection{Phase F2}
 To understand this phase, we start from the $t'=0$ case, where the
 Hamiltonian (\ref{ferroh}) is proportional to
$2 \sum_{\rm bonds} \left({\bf n}_{ij}^z {\bf T}_i \right) 
 \left( {\bf n}_{ij}^z{\bf T}_j\right)$, which can conveniently be 
transformed to 
\begin{eqnarray*}
&&\sum_{\rm bonds}
\left(  \left[ {\bf n}_{ij}^z \left({\bf T}_i +{\bf T}_j\right)\right]^2 - 
  \left({\bf n}_{ij}^z {\bf T}_i \right)^2 -
                   \left({\bf n}_{ij}^z {\bf T}_j \right)^2
\right)
\nonumber  \\
 &=&\sum_{\rm bonds}
\left[{\bf n}_{ij}^z \left({\bf T}_i +{\bf T}_j\right) \right]^2 - 3 
\sum_i 
  \left[ \left(T^x_i\right)^2 + \left(T^z_i\right)^2 \right]
\nonumber  \\
 &=& \sum_{\rm bonds}
\left[ {\bf n}_{ij}^z\left({\bf T}_i +{\bf T}_j\right)\right]^2 
+ 3 
\sum_i \left(T^y_i\right)^2 
- 3 N T(T\!+\!1) \nonumber  \\
\end{eqnarray*} 
At the classical level, the two squares can be minimized by choosing the 
${\bf T}$ vector in the $(T^x,T^z)$ plane so that on a given bond either 
${\bf T}_i=-{\bf T}_j$, or ${\bf T}_i+{\bf T}_j$ is perpendicular to 
${\bf n}_{ij}^z$.
These conditions are satisfied with the collinear orbital order 
 shown in Fig.~\ref{fig:tabfm}: we choose $|a\rangle + |b\rangle$ along 
every second chain with the bond variable ${\bf n}_{ij}^z=(0,0,1)$, and 
$|a\rangle - |b\rangle$ along the remaining chains (the orbital configuration
is the same as in phase B in Fig.~\ref{phases}). There are 6 such 
configurations, 
which can be obtained by translations and rotations, with variational energy 
\begin{equation}
  \frac{E_{\rm CL}}{N} = -\frac{1}{4}\frac{3t^2-2tt'+3t'^2}{\tilde U - J_H}
\end{equation}
The classical phase boundaries for this state are $t'/t=-1/3$ and $t'/t=1/3$.

\begin{figure}[ht]
\begin{center}
\includegraphics[width=7truecm]{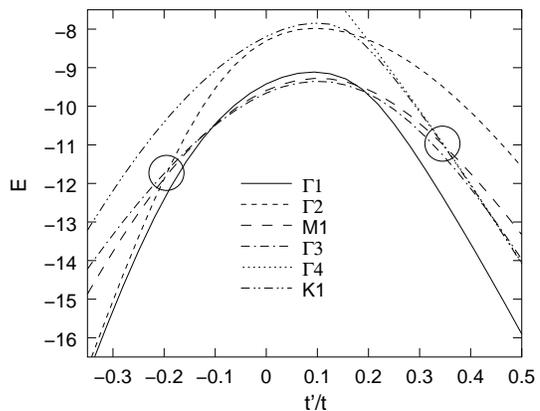}
\caption[sf]{\label{fig:FM_ED} 
  Energy level scheme of a 12 site diamond-like cluster with periodic boundary
  conditions and compatible with the 3-sublattice LRO. 
Shown are the levels 
which can be associated with the
  ferro-orbital $T^y$ order (denoted by $\Gamma1$ and $\Gamma2$), collinear
  phase ($\Gamma1$, $M1$, and $\Gamma3$) and the lowest states constituting
  the Anderson tower of the 120$\,^{\circ}$ antiferro-orbital phase ($\Gamma1$,
  $K1$, and $\Gamma4$). The first letter refers to the momentum of the state.
  We have encircled the level crossings which we used to determine the phase
  boundary ($t'/t=-0.20$ and $t'/t=0.35$).  }
\end{center}
\end{figure}

In a finite system with
periodic boundary conditions respecting the point group ${\cal D}_{3d}$ of the
trianguar lattice, the linear combination of the 6 states will produce a
3-fold degenerate state at the $M$ point in the Brillouin zone 
(state $M1$ in Fig.~\ref{fig:FM_ED}), and 3 states at the $\Gamma$ point,
one non-degenerate and one 2-fold degenerate
($\Gamma1$, and $\Gamma3$ in
Fig.~\ref{fig:FM_ED}, respectively). These states can clearly be
recognized in the exact diagonalization spectrum of the 12 site cluster as the
lowest lying state for $-0.2t<t'<0.35t$, well separated from the states with
higher energy. The observation of the phase in the correlation function is
non-trivial, as the ground state around $t'=0$ is twofold degenerate, and the
applied exact diagonalization on a finite size cluster will result in a state
with an  arbitrary linear combination of them, which leads to a 
pattern difficult to interpret. It is, however, clear 
that there is no ferro-orbital order.

\subsubsection{Phase F3}
In this phase the  $T^y$ ferro-orbital order is established: for negative
$t'$ the ${\bf T}_i {\bf T}_j$ term in Eq.~(\ref{ferroh}) becomes 
ferromagnetic, and the frustration in the $T^x$ and $T^z$ due to the 
$({\bf n}_{ij}^z{\bf T}_i) ({\bf n}_{ij}^z{\bf T}_j)$ term
will single out the $T^y$ order. The particularity of the $T^y$ ordering is that it breaks the
time-reversal symmetry: either the $|a\rangle + i|b\rangle$ or the  
$|a\rangle - i|b\rangle$ combination orders.    
The ordering of complex orbitals has been searched for in the context of
manganites, where it has been thought that they are favoured by the isotropic
kinetic exchange. Indeed, the charge density of the $|a\rangle \pm i|b\rangle$
shows the trigonal symmetry, and the combination is Jahn-Teller inactive. 
The phase can be easily identified in the finite size diagolization from 
 the correlation function: spatially isotropic $T^y_i T^y_j>0$ correlations are
 dominant. The mean field variational energy of the
 ferro-orbital complex state is 
\begin{equation}
  \frac{E_{\rm FO}}{N} = \frac{3tt'}{\tilde U - J_H}
\end{equation}
and the phase is stable for $t'/t<-1/3$.

 The determination of the phase boundaries is, however, not straightforward.
As can be seen from the
energy levels, the $\Gamma1$ state is present in the `ground-state manifold'
of all the ordered phases. Therefore we identified the phase boundaries by
level crossings of the `ground-state manifolds' associated with each type of
ordering, which agree reasonably well with the classical phase boundaries
($t'/t=\pm 1/3$). 
At these phase boundaries continuous degeneracies
appear in the classical wave function, suggesting a gapless excitation 
spectrum at those points. 

\section{Exact diagonalizations}

\begin{figure}[ht]
\begin{center}
\includegraphics*[width=4truecm,angle=0]{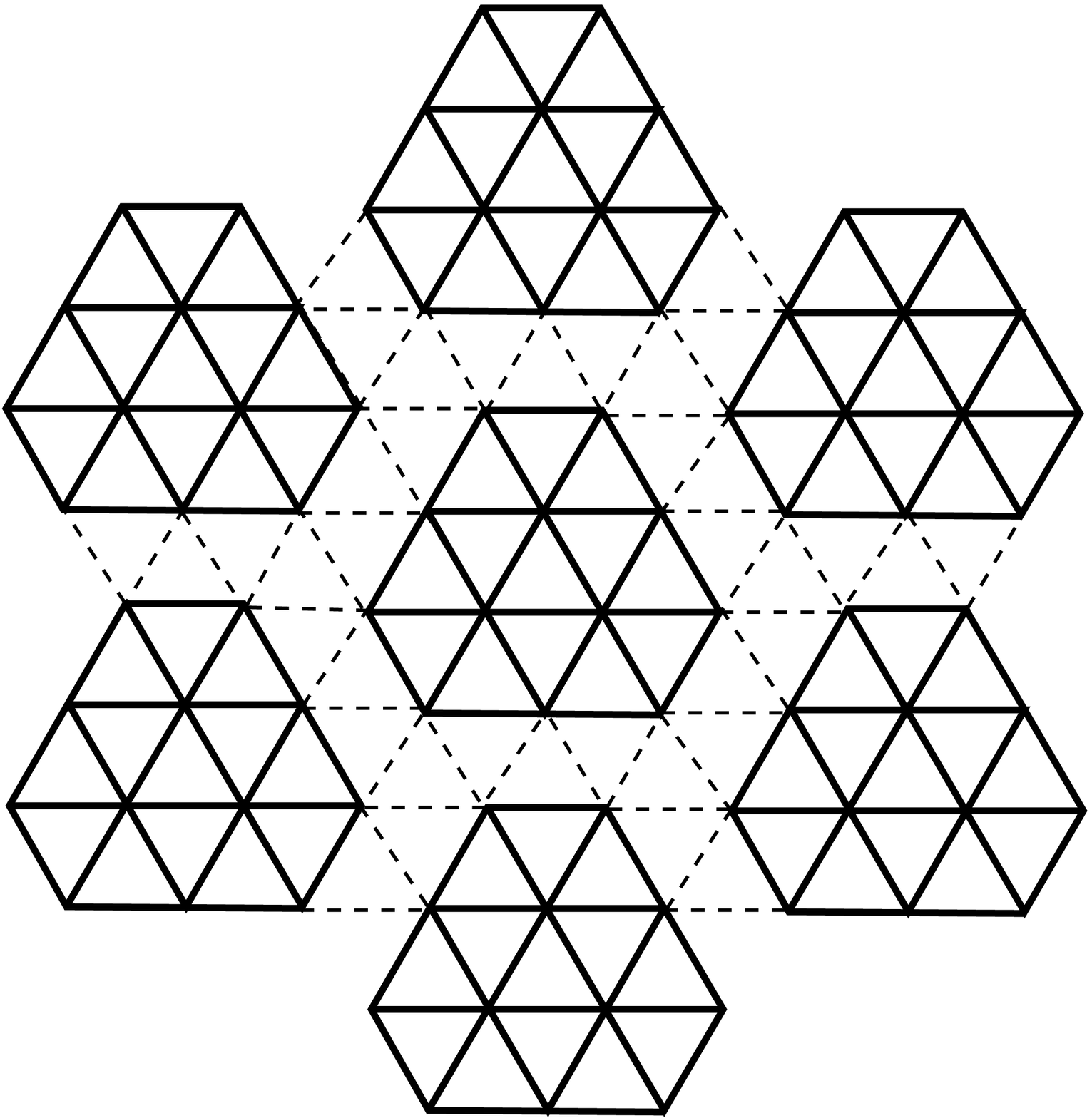}\\
\includegraphics*[width=4truecm,angle=0]{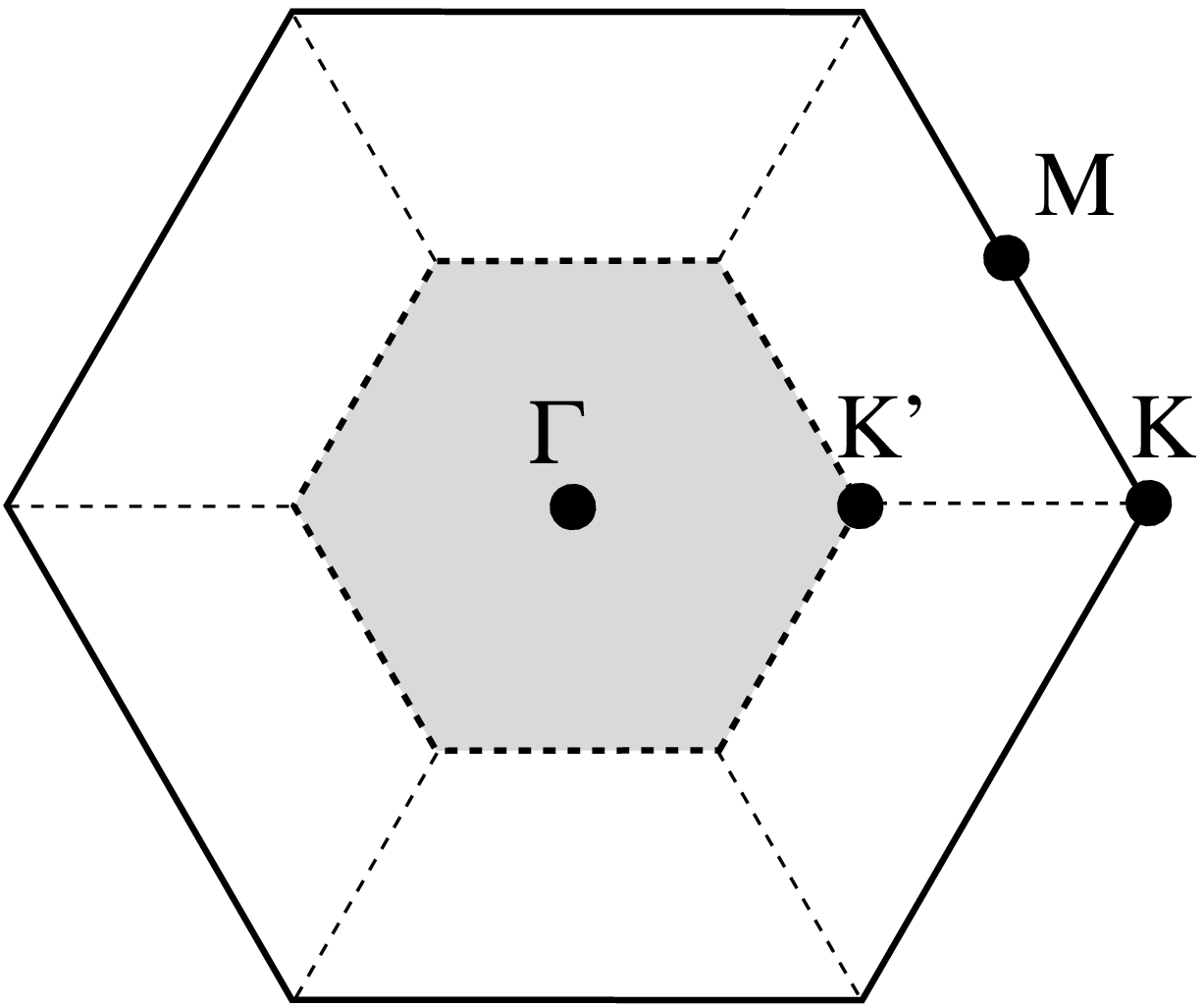}
\caption[clust]{\label{clust}{The 12-site cluster with periodic boundary 
conditions, and the associated Brillouin zone}}
\end{center}
\end{figure}

 Due to the small number of conveniently exploitable symmetries 
in the problem (we have only 
the spin SU(2) symmetry), the size of the Hilbert space grows very rapidly
with the size. 
In the $S^z=0$ sector it increases like ${N/2\choose N}2^N $ where 
$N$ is the number of sites. This limits us to small cluster sizes,
especially if we want to explore the phase diagram. 
The obvious choice was the  
12-site cluster with periodic boundary conditions (Fig.~\ref{clust}), which 
has the full point ${\cal D}_{3d}$ symmetry of the lattice as well. The considered 
cluster has the advantage to allow the formation of SU(4)  plaquettes, 
and is also compatible with three and four sublattice order.

\begin{figure}[ht]
\begin{center}
\includegraphics*[width=7truecm,angle=0]{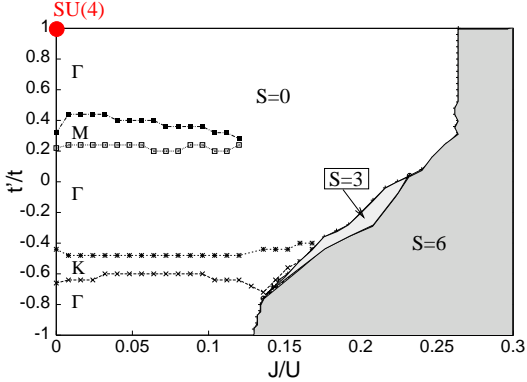}
\caption[var-dia]{\label{exact}
Exact diagonalizations : phase diagram for the 12-site cluster
}
\end{center}
\end{figure}

The phase diagram obtained from the level crossings in the ground state
 is shown in Fig.~\ref{exact}. It is globally
consistent  with the mean-field one. The fully polarized ferromagnetic 
region ($S=6$) is found for very similar values of $J/U$. For small
$J/U$, we identify 5 different regimes from $t'/t=-1$ to
$t'/t=+1$. They seem to correspond to 4 phases only since two
regions join for intermediate values of $J/U$, but given the 
difficulty to determine phase boundaries from exact diagonalizations,
this should not be taken too seriously.
The various regions are labelled according to the point in the Brillouin 
zone where the ground state is found. 
 
In the vicinity of the SU(4)-point ($t'/t\sim1$,$J\sim0$), the low-lying 
spectrum is similar to the one obtained in the SU(4) case\cite{su4penc}. 
This suggests 
that the description of the ground state in terms of SU(4) singlets 
is applicable in this region.

At the $M$ point the ground state is three-fold degenerate. This could
correspond to the  formation of AF chains in phase B since these
chains can form in three directions, resulting in a 3-fold degenerate
mean-field solution.  To confirm this interpretation, we have diagonalized
the full Hamiltonian in the variational sub-space spanned by the mean-field
ground state wave -functions of Phase B. It turns out that these states 
are not coupled
because, due to the orbital configuration, they have different symmetries
with respect to the inversion around the middle points of nearest-neighbour
bonds. So the ground-state degeneracy in this variational subspace is 
still equal to 3, supporting the interpretation in terms of chains.

When the ground-state is at the $K$ point,
the interpretation is not so straightforward. The ground state is strictly 
speaking twofold degenerate. But looking at the spectra the first excited 
state is at the $\Gamma$ point, and very close to the ground state. 
A possible explanation could be that all these states are degenerate in the 
thermodynamic limit.Then this region could also be explained by the formation 
of chains. To check this point, we have diagonalized the Hamiltonian
in the variational sub-space spanned by the three mean-field
ground state wave -functions of Phase D. The orbital configuration is different
from Phase B, and the degeneracy is partially lifted, with a two-fold
degenerate ground-state an a non-degenerate excited state. Again the
agreement supports the interpretation of this phase in terms of AF chains.

\begin{figure}[ht]
\begin{center}
\includegraphics*[width=5truecm,angle=-90]{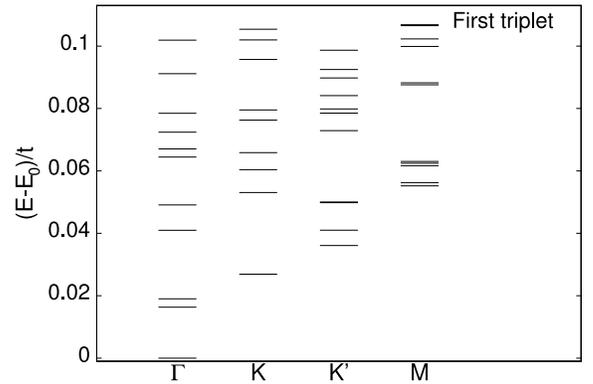}
\caption[spec]{\label{spec}
 Low-lying singlets for $t'/t=0$ and $J/U=0.008$ }
\end{center}
\end{figure}

The most interesting region for our case is the central one. 
We will focus our attention on the line $t'/t=0$. Along this line we will see 
that a description  in terms of RVB states is reasonable.
For instance, the low-lying spectrum (Fig.~\ref{spec})
 for $J/U=0.008$ and $t'/t=0$, 
shows a very large number of singlets states (125) before the first 
triplet (at the top of the figure). All these singlets are very close 
in energy, the energy difference between the ground 
state and the first triplet being of the order of $\sim\frac{t^2}{U}$.
Note that the number of singlets below the first triplet (125) is a 
significant fraction of the total number of dimer-coverings for 
this 12-site cluster (348). This is reminiscent of the spectrum found
by Lecheminant et al for the S=1/2 Heisenberg model on the {\it kagome} 
lattice\cite{lecheminant}.

\begin{figure}[ht]
\begin{center}
\includegraphics*[width=7truecm,angle=0]{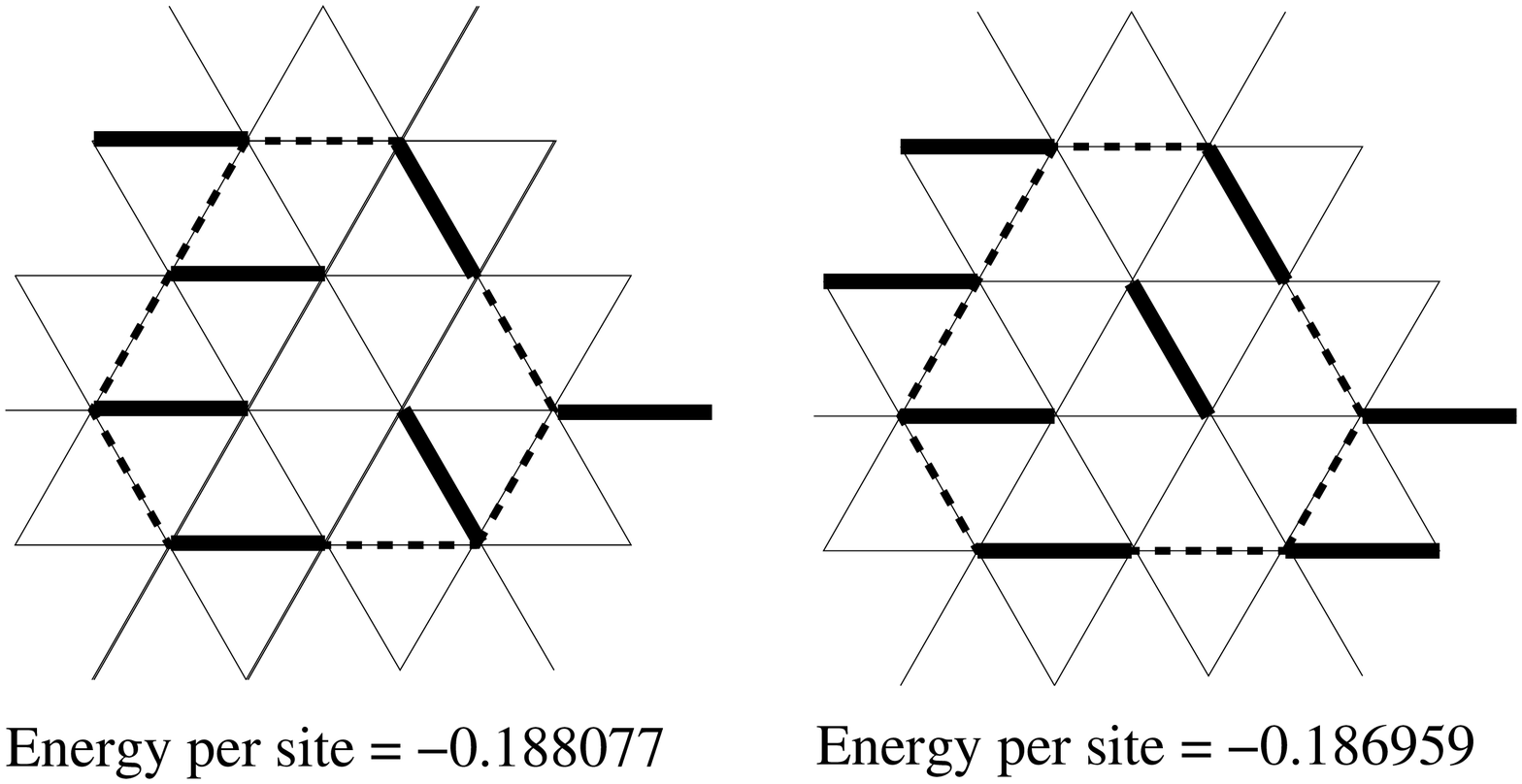}
\caption[dimpat]{\label{dimpat}
Two stable states for a 12-site 
cluster : the dashed line represents the cluster. $t'/t=0$ and $J/U=0.008$
}
\end{center}
\end{figure}

This is in qualitative agreement with the mean-field results.
Indeed, in phases C and C', several solutions corresponding to
various dimer coverings were found with comparable energies
(see Fig.~\ref{dimpat}). A similar observation was made in a preliminary  
study of a similar spin--orbital model in the context of
BaVS$_3$\cite{bavs}.

A possible ground state for this region could be a spin--orbital version of 
the resonating valence bond 
(RVB) state\cite{rvb}. The magnetic structure could be envisaged as a 
fluctuating pattern of bonds among different dimer-coverings or a 
mixture between dimer-coverings and chains. All these states being singlets,  
they may be degenerate in the thermodynamic limit.

Let us also mention that there is also a partially polarized region $S=3$. 
We suspect that it may be a finite-size effect, as it is greatly reduced 
with respect to the corresponding $S=1$ phase  present in 
the phase diagram of the tetrahedron (Fig.~\ref{fig:diag4}).

\begin{figure}[ht]
\begin{center}
\includegraphics*[width=7truecm,angle=0]{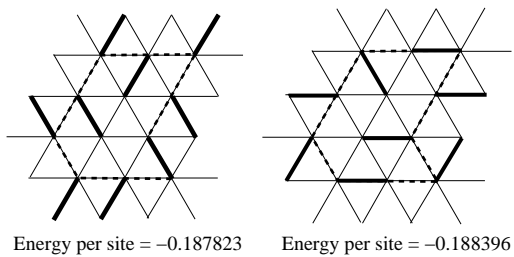}
\caption[dimpat]{\label{dimpatbis}
Two stable states for another 12-site 
cluster : the dashed line represents the cluster. $t'/t=0$ and $J/U=0.008$}
\end{center}
\end{figure}   
\section{Experimental implications}

In transition metal oxides, the on-site Coulomb repulsion $U$ is
typically in the range 4-10 eV, and the Hund's rule coupling
in the range 0.5-1 eV, leading to a physical range defined
by $0.05 < J/U < 0.25$. Interestingly enough, all phases appear
in this range and should be possible to observe in actual
compounds provided the
ratio $t'/t$ has the appropriate value. 
In that respect, one should emphasize that the phase 
diagram depends only on the hopping integrals between orbitals, 
not on the actual orbitals.
In particular, even if the two orbitals were not orthogonal
by symmetry on one of the bonds, diagonalizing the hopping matrix
on a given bond would bring us back to the situation treated
in this paper. So the discussion would carry over beyond the specific
case of $d_{3z^2-r^2}$ and $d_{x^2-y^2}$ up to the remarks
dealing with Jahn-Teller distortions.

Now, coming back to LiNiO$_2$ and NaNiO$_2$, hence to $d_{3z^2-r^2}$ 
and $d_{x^2-y^2}$ orbitals, simple arguments suggest
that $t'/t$ is {\it negative} and {\it small}. That it is negative 
comes from the different symmetries of the orbitals: the $d_{3z^2-r^2}$
orbitals on edge-sharing octahedra (see Fig. 2) are symmetric with 
respect to the mirror plane that brings one octahedron into the
other, while the $d_{x^2-y^2}$ are antisymmetric. Now any direct
overlap between $d$ wave functions in transition metal oxides is
known to be very small. However one should not forget that the 
orbitals are in fact Wannier functions centered on the transition
metal ions which extend in general to infinity to insure orthogonality,
and which have a significant weight on neighbouring O $2p$ orbitals.
In the case of the Wannier orbitals with symmetry $d_{x^2-y^2}$,
this does not lead to any significant transfer because the O $2p$ 
orbitals coupled to one of them are orthogonal to the $d_{x^2-y^2}$
of the neighbouring octahedron. This is not strictly true here 
since the Ni-O-Ni angle is not exactly 90$\,^{\circ}$, and also because
the crystal field is not symmetric at the oxygen site, but still
one expects the effective hopping to be very small. By contrast, the 
$d_{3z^2-r^2}$ Wannier orbitals have weight on the O $2p$ orbitals
above and below, and these O $2p$ orbitals have a standard $\pi$ 
overlap regardless of the actual local distortions of the octahedra.
So this should give rise to a significant overlap between the 
Wannier functions with $d_{3z^2-r^2}$ symmetry. 

Beyond the actual value of the parameters, it is important to 
emphasize that we have not adopted the same point of view as Mostovoy
and Khomskii~\cite{most}, who have neglected any overlap
between Ni orbitals, although it is allowed by symmetry. Further, they have
assumed that Ni-O-Ni bonds make and angle of 90$\,^{\circ}$, although the actual
angle is around 94$\,^{\circ}$ in LiNiO$_2$ and 96.4$\,^{\circ}$ in 
NaNiO$_2$, and they have neglected the role of crystal
field at the oxygen site, known to produce antiferromagnetic 
couplings as shown by Dar\'e et al~\cite{dare}. While the ferromagnetic 
coupling
that comes out of these approximations is certainly relevant, the
simplified Hamiltonian studied by Mostovoy and Khomskii leads
to a purely ferromagnetic coupling, while the more general Hamiltonian
studied in the present work exhibits a rich variety of phases
which, we believe, might actually lead to the ultimate explanation of 
LiNiO$_2$ and NaNiO$_2$.

\subsection{LiNiO$_2$}

In the case of LiNiO$_2$, which undergoes neither a Jahn-Teller
distortion nor a magnetic phase transition upon lowering the 
temperature, 
 we have to choose between two different realizations of RVB: the SU(4) phase
A (for $t\approx t'$), and the fluctuating dimer phases C and C'. 
Since we have argued that $|t'/t| \ll 1$, 
we opt for the dimer phases.
Actually, one should give 
preference to Phase C' since $t'/t$ is negative, but
as we discussed these phases should better be considered as
defining a domain in which the physics of the quantum
dimer model (QDM) on the triangular lattice might be relevant.
The actual form of the effective QDM is not known yet, but it
presumably will not be too far from the minimal model studied
by Moessner and Sondhi~\cite{moessner} since phase C is a staggered state
and belongs to the ground state manifold of their model
for large engouh repulsion between face-to-face dimers,
while phase C' is a maximally flippable state and belongs
to the ground state manifold in the limit of infinite
attraction between face-to-face dimers. Note however that
Phase C' is {\it not} the columnar state realized for finite
attraction in the minimal model, so differences are to be
expected. Still, close to the boundaries between the phases,
one may speculate that an RVB phase will be present. Such
a phase does not break any symmetry and could explain
the absence of any kind of ordering in LiNiO$_2$.
Let us also note that the EXAFS results by Rougier et al~\cite{rougier}
are also consistent with this proposal since the orbitals entering
all these states are Jahn-Teller orbitals with two long
bonds and four short bonds. If the system undergoes resonances
between different states, this would produce a dynamic
Jahn-Teller effect between these states, a situation still
leading to two long bonds and four short bonds on average.
Due to some disorder, and/or to coupling to the lattice,
the system might actually prefer to freeze in a non-periodic
dimer covering of the triangular lattice, as suggested by 
Reynaud et al~\cite{rey}. Such a frozen, non-periodic state would also be
consistent with the results of Ref.[~\onlinecite{rougier}].

\subsection{NaNiO$_2$}

As far as NaNiO$_2$ is concerned, the only potential candidate
is phase D since this is the only ferro-orbital phase with 
Jahn-Teller orbitals consistent with the distortion that 
occurs at 480 K in that system. This phase has the largest
boundary with phase C', a good point in view of the very 
similar structures of LiNiO$_2$ and NaNiO$_2$. 
As stated earlier, the effective model consists of weakly coupled AF chains,
and the resulting magnetic structure will depend on the residual
couplings. A thorough analysis of this point will require to go
beyond the present calculation and is left for future
investigation. But in any case, with some interlayer coupling,
this is expected to lead to some kind of AF ordering at finite temperature,
in agreement with experiments. 
Let us emphasize that, while simultaneous ferromagnetism {\it and} 
Jahn-Teller active ferro-orbital
order have been argued to be possible by Mostovoy and 
Khomskii~\cite{most} in the context of their simplified model, 
this seems to be impossible in the context of our microscopic model.
Now, as far as experiments are concerned, the actual order
is not known yet. It has been often assumed so far that
this AF state consists of ferromagnetic planes coupled
antiferromagnetically, but new, preliminary results seem
to indicate that this cannot be the case~\cite{chouteau}, which
opens the way for another type of antiferromagnet. 

\subsection{Curie-Weiss constant}

\begin{figure}[ht]
\begin{center}
\includegraphics*[width=7truecm,angle=0]{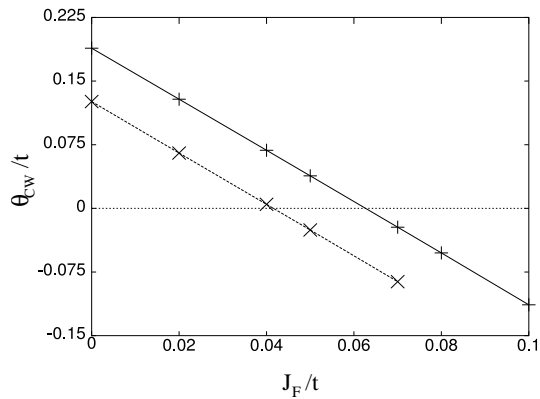}
\caption[HvsJF]{\label{HvsT}{$\theta_{CW}$ as a function of $J_F$ 
for C and D phases (see fig.~\ref{var-dia}): $J/U=0.064$ and $t'/t=-0.1$ for 
the solid line (D phase), $t'/t=0.1$ for the dashed line (C phase)}}
\end{center}
\end{figure}

At this stage, the essential problem when comparing our 
predictions to the experimental data for LiNiO$_2$ and 
NaNiO$_2$\cite{chap-mag} is the sign of the Curie-Weiss constant,
which is a measure of the average coupling. In both
cases, it is ferromagnetic if determined at not too high
temperature\cite{rey}, while in our calculation,
it is antiferromagnetic. This is not a very serious problem
however. In deriving our model, we have only kept second
order terms in the hopping $t$ and $t'$ between Wannier
orbitals centered at Ni sites. This derivation, exactly 
similar to the derivation of kinetic exchange, neglects
the ferromagnetic coupling that is always present due
to the overlap of these Wannier functions. 
The precise form of this ferromagnetic coupling in the context 
of extended Wannier functions centered at the nickel site, which is 
the point of view adopted in the present paper, has not been derived yet 
but it is expected to be similar to the form derived by Mostovoy and Khomskii.
In any case, we have checked that if we include an {\it ad hoc}
ferromagnetic term 
 (the B' term in Eq.~(\ref{eq:effham3ph}))
\begin{equation}
 H=-J_F \sum_{i,j}{\bf S}_i.{\bf S}_j
\end{equation}
to the Hamiltonian ($J_F>0$), the phases discussed above
remain stable in a region where the Curie-Weiss constant
is ferromagnetic. 
More precisely, we have solved the self-consistent equations including such
a term, which leads to the effective spin Hamiltonian: 
\begin{equation}
 H=\sum_{i,j}\left\{({\bf S}_i.{\bf S}_j)h'_{ij}+k_{ij}^T\right\}
\end{equation}
where $h'_{ij}=\langle h_{ij}^T \rangle -J_F$, and the Curie-Weiss
constant is given by:
\begin{equation}
\theta_{CW}=\frac{S(S+1)}{3}\sum_{j(i)}h'_{ij}=\frac{1}{2}{\sum_{<ij>}}' h'_{ij}
\end{equation}
where $\sum_{j(i)}$ means summation over all first neighbours of a a given
site $i$, while $\sum'_{<ij>}$ means summing over
three pairs of nearest-neighbours in three inequivalent directions.
The results are summarized in figure \ref{HvsT}. As announced, the 
Curie-Weiss temperature changes sign inside phases C and D before
one enters the ferromagnetic phase, and this occurs for values
of $J_F$ which are small enough to be physically relevant.

\section{Conclusion}
We have shown that a spin-orbital model on the triangular lattice with 
realistic parameters leads to a very rich physics. 
The presence of various important phases (SU(4), dimers and ferromagnetic) is 
confirmed for every cluster (4, 12 and 16 sites).
Moreover it seems that 
this model is able to provide a good description of the behaviour of LiNiO$_2$ 
and 
NaNiO$_2$, and to explain the puzzling difference between these two compounds. 
We have given 
 specific
  meaning to the claim that an RVB 
state seems to be at the origin of the magnetic properties of LiNiO$_2$. The 
underlying orbital structure corresponding to this RVB state is in 
agreement with the experimental observations. 
For the case of NaNiO$_2$, a new possible magnetic state has been investigated 
with an  underlying orbital structure which still leads to a cooperative 
Jahn-Teller distortion.
A precise description of the low energy physics of the present model for the 
phases relevant for LiNiO$_2$ and NaNiO$_2$ requires other methods than those 
used in the present paper, but we are confident that the present 
analysis will set the stage for further investigations.

\begin{acknowledgments}
F.V. wishes to acknowledge M. Mambrini and F. Becca for their useful help.
K.P. and P.F. were supported by the Hungarian national grants OTKA T 038162, 
T 037451, and D32689. F. M. and F. V. acknowledge support from the 
Swiss National Fund.
\end{acknowledgments}


\begin{thebibliography}{99}

\bibitem{most} M.V. Mostovoy and D.I. Khomskii: Phys. Rev. Lett. 
{\bf 89}, 227203 (2002). 

\bibitem{chap-JT} E. Chappel, M. D. Nunez-Regueiro, G. Chouteau, O. Isnard,
  and C. Darie, Eur. Phys. J. B {\bf 17}, 615 (2000).

\bibitem{yamaura} K. Yamaura, M. Takano, A. Hirano, and R. Kanno:
    J. Solid State Chem. {\bf 127}, 109 (1996).

\bibitem{otherE} Specifically, we mean the $E$ derived from $e_g$. It will mix 
with the $E$ derived from $t_{2g}$, but this effect does not appear
explicitely in our reasoning.

\bibitem{chap-mag} E. Chappel, M.D. Nunez-Regueiro, F. Dupont, G. Chouteau, 
C. Darie, A. Sulpice, Eur. Phys. J. B {\bf 17}, 609 (2000).

\bibitem{rougier} At least not to an identifiable long-range pattern. EXAFS 
studies (A. Rougier, C. Delmas, and A.V. Chadwick: Solid State 
Commun. {\bf 94}, 123 (1995)) indicate that there is a local distortion at 
every Ni site, but the distortion pattern is frustrated, and does not support 
an (orbital) ordering transition.

\bibitem{hirota} K. Hirota, Y. Nakazawa, and M. Ishikawa, J. Phys.: Condens. 
Matter {\bf 3}, 4721 (1991).

\bibitem{kitaoka} Y. Kitaoka et al., J. Phys. Soc. Jpn. {\bf 67}, 3703
    (1998).


\bibitem{olesrev} For a review, cf. 
A.M. Ole\'s, M. Cuoco, and N.B. Perkins, {\sl in}: {\sl Lectures on
the Physics of Highly Correlated Electron Systems}, Ed. F. Mancini, AIP
Conf. Proc. {\bf 527}, p. 226 (New York, 2000).

\bibitem{rvb} P.W. Anderson: Mater. Res. Bull. {\bf 8}, 153 (1973).

\bibitem{kleine} P. Fazekas and P.W. Anderson: Phil. Mag. {\bf 30}, 423 
(1974); B. Kleine, E. M\"uller-Hartmann, K. Frahm, and P. Fazekas: 
Z. Phys. B: Condens. Matter {\bf 87}, 103 (1992).

\bibitem{rey} F. Reynaud et al, Phys. Rev. Lett. {\bf 86}, 3683 (2001). 

\bibitem{hira} K. Hirakawa, H. Kadowaki, and K. Ubukoshi, J. Phys. Soc. 
Japan {\bf 54}, 3526 (1985).

\bibitem{arim} T. Arimori and S. Miyashita: J. Phys. Soc. Japan {\bf 69}, 
2250 (2000).

\bibitem{su4penc} K. Penc, M. Mambrini, P. Fazekas, and F. Mila: Phys. Rev. B 
{\bf 68}, 012408 (2003).

\bibitem{KK} Kugel and Khomskii : Sov. Phys. Usp. {\bf 25}, 232, (1982).

\bibitem{cast} C. Castellani, C.R. Natoli, and J. Ranninger: Phys. Rev. 
B {\bf 18}, 4945; 4967; 5001 (1978). 

\bibitem{double} Spin-orbital states should be classified according to the 
double group belongig to ${\cal D}_{3d}$. This would be important if we 
included the effect of the relativistic spin-orbit interaction. In the present 
work, we assume that the spin and orbital Hilbert spaces are decoupled, and 
total spin eigenstates can be combined with point group basis states.

\bibitem{inver} We could have chosen the inversion instead of $\sigma_h$.

\bibitem{lattice} For pairs of different orientation, or situated elsewhere in 
the lattice, ${\cal C}_{2}$ or/and $\sigma_h$ should have been chosen 
differently. 
Summing over all pairs, the resulting hamiltonian respects 
the space group symmetry of the lattice.

\bibitem{dare} A.M. Dar\'e, R. Hayn, and J.L. Richard: Europhys. Lett. 
{\bf 61},  803 (2003).

\bibitem{darenote} Apart from a remark that the effect of $J_{\rm H}$ is 
insignificant. Our phase diagrams (Figs.~\protect{\ref{var-dia},\ref{exact}}) show 
that new phases are found at $J_{\rm H}\ne 0$.

\bibitem{fczhang} Y. Q. Li, M. Ma, D. N. Shi, and F. C. Zhang:
  Phys. Rev. Lett. {\bf 81}, 3527 (1998).

\bibitem{ops} This choice of order parameters is permissible in the absence of
  spin--orbit coupling. With the spin--orbit coupling included, the right 
choice would be to use operators which are basis elements for the irreps of
the ${\cal D}_{3d}$ double group.

\bibitem{bavs} G. Mih\'aly, I. K\'ezsm\'arki, F. Z\'amborszky, M. Miljak, 
K. Penc, P. Fazekas, H. Berger, and L. Forr\'o: Phys. Rev. B{\bf 61}, 
R7831 (2000).

\bibitem{afm} In the sense that the fluctuating antiparallel correlations of 
a singlet are extrapolated to the antiparallel correlations of quasi-classical 
antiferromagnetic order. For spin--orbital models, the reasoning cannot be 
so straightforward.

\bibitem{typol} We note that the cubic $e_g$ doublet system has a 
Hartree--Fock instability against $T^y$-polarization at non-integral 
fillings, though this solution does not seem motivated by the 
spin--orbital hamiltonian (A. Takahashi and H. Shiba, 
J. Phys. Soc. Jpn. {\bf 69}, 3328 (2000)).

\bibitem{others} For a pair of different orientation, it would be axial 
symmetry about a different direction in pseudospin space. It is for this 
reason that axial symmetry of the pair cannot be restated as a global 
symmetry of the lattice.

\bibitem{notethat} Note that this section of the line is not a phase 
boundary, but it lies inside the ($S=1$, $T=0$) phase. Selecting the 
non-degenerate state $|ab\rangle-|ba\rangle$ does not require any special 
orbital symmetry.

\bibitem{frederic} F. Mila, Phys. Rev. Lett. {\bf 81}, 2356 (1998).

\bibitem{su4but} We are not too worried about this point for the following 
reason: the tetrahedral symmetry is atypically high, and probably gives a 
too large preference to SU(4) solutions. $N>4$ pieces of the triangular 
lattice (even with periodic boundary conditions imposed) are less 
symmetrical.  

\bibitem{frischmuth} This was clearly established for the 1D case, where
the energy is dominated by the mean-value of the four operator term.
See e.g. B. Frischmuth, F. Mila, M. Troyer, Phys. Rev. Lett. {\bf 82}, 
835 (1999).

\bibitem{coldea} See e.g. R. Coldea et al, Phys. Rev. Lett. {\bf 88}, 137203 (2002)
and references therein.

\bibitem{bernu} B. Bernu, C. Lhuillier, and L. Pierre, Phys. Rev. Lett.
{\bf 69}, 2590 (1992).

\bibitem{lecheminant} P. Lecheminant et al, Phys. Rev. B {\bf 56}, 2521 (1997).

\bibitem{moessner} R. Moessner and S. L. Sondhi, Phys. Rev. Lett. {\bf 86}, 
1881 (2001).

\bibitem{chouteau} G. Chouteau and S. De Brion, private communication.

\end{thebibliography}
\end{document}